\documentclass[12pt]{article}
\usepackage{latexsym}
\usepackage{amsmath}
\usepackage{amsfonts}
\usepackage{amssymb}
\usepackage{amscd}
\usepackage{fancybox}
\usepackage{cite}
\usepackage{amsmath,amsfonts,amsbsy}
\usepackage{pstricks,pst-node}
\usepackage{graphicx}
\usepackage{epsfig}
\usepackage{psfrag}

\DeclareMathOperator{\extdm}{d}
\newcommand{\extd}{\extdm \!}

\usepackage{float}

\psset{unit=1.3cm,linewidth=.5pt,radius=.2}  

\usepackage{float}                          
\usepackage{lscape}                         
\usepackage{bm}

\newcommand{\cF}{{\cal F}}
\newcommand{\cG}{{\cal G}}
\newcommand{\cK}{{\cal K}}
\newcommand{\cL}{{\cal L}}
\newcommand{\cM}{{\cal M}}

\newcommand{\cO}{{\cal O}}

\newcommand{\beq}{\begin{equation}}
\newcommand{\eeq}{\end{equation}}
\newcommand{\bi}{\begin{itemize}}
\newcommand{\ei}{\end{itemize}}
\newcommand{\bt}{\begin{tabular}}
\newcommand{\et}{\end{tabular}}
\newcommand{\bc}{\begin{center}}
\newcommand{\ec}{\end{center}}

\newcommand{\ket}[1]{|#1\rangle}

\newcommand{\bracket}[2]{\langle#1|#2\rangle}
\newcommand{\vev}[1]{\langle#1\rangle}

\newcommand{\be}{\begin{equation}}
\newcommand{\ee}{\end{equation}}
\newcommand{\bea}{\begin{eqnarray}}
\newcommand{\eea}{\end{eqnarray}}
\newcommand{\ba}{\begin{array}}
\newcommand{\ea}{\end{array}}

\def\bbox{{\,\lower0.9pt\vbox{\hrule \hbox{\vrule height 0.2 cm
\hskip 0.2 cm \vrule height 0.2 cm}\hrule}\,}}
\newcommand{\dsl}{\pa \kern-0.5em /}

\newcommand{\rmd}{\mathrm{d}}

\newcommand{\bg}{\bar g}

\newcommand{\R}{\bar R}

\newcommand{\bBox}{\bar{\square}}
\newcommand{\bnabla}{\bar{\nabla}}

\begin{document}

\begin{titlepage}
\begin{center}

\hfill UG-12-77

\vskip 1.5cm

{\large \bf Three-Dimensional Tricritical Gravity  
}

\vskip 1cm

{\bf Eric A.~Bergshoeff, Sjoerd de Haan, Wout Merbis,\\ [.3truecm]
 Jan Rosseel and  Thomas Zojer}

\vskip 25pt

{\em\hskip -.1truecm Centre for Theoretical Physics,
University of Groningen, \\ Nijenborgh 4, 9747 AG Groningen, The
Netherlands \vskip 5pt }

{email: {\tt E.A.Bergshoeff@rug.nl, s.de.haan@rug.nl, w.merbis@rug.nl, j.rosseel@rug.nl, t.zojer@rug.nl}} \\
\vskip 15pt

\end{center}

\vskip 1.5cm

\begin{center} {\bf ABSTRACT}\\[3ex]
\end{center}

We consider a class of parity even, six-derivative gravity theories in three dimensions.
After linearizing around AdS, the theories have one massless and two massive graviton
solutions for generic values of the parameters.
At a special, so-called tricritical, point in parameter space the two massive graviton solutions become
massless and they are replaced by two solutions with logarithmic and logarithmic-squared boundary behavior. The theory at
this point is conjectured to be dual to a rank-3 Logarithmic Conformal Field Theory (LCFT) whose boundary
stress tensor, central charges and new anomaly we calculate. We also calculate the conserved Abbott--Deser--Tekin charges.
At the tricritical point, these vanish for excitations that obey Brown--Henneaux and logarithmic boundary conditions,
while they are generically non-zero for excitations that show logarithmic-squared boundary behavior. This suggests that
a truncation of the tricritical gravity theory and its corresponding dual LCFT can be realized either via boundary
conditions on the allowed gravitational excitations, or via restriction to a zero charge sub-sector. We comment on the
structure of the truncated theory.

\end{titlepage}

\newpage
\setcounter{page}{1} \tableofcontents
%

\section{Introduction}

Higher-derivative theories of gravity in $d \geq 3$ have recently received a lot of attention. In three dimensions,
higher-derivative theories have been used to construct models that allow for the propagation of massive bulk
gravitons, thus leading to non-trivial models of three-dimensional (massive) gravity. Examples are Topologically
Massive Gravity (TMG) \cite{Deser:1981wh} and New Massive Gravity (NMG) \cite{Bergshoeff:2009hq}, which have
resp.~third and fourth order derivative terms. The combination of TMG and NMG leads to so-called General Massive Gravity
(GMG), that is the most general three-dimensional gravity model with up to four derivatives that propagates only
spin-2 excitations \cite{Bergshoeff:2009hq,Bergshoeff:2009aq}. All these models contain a number of dimensionless
and dimensionfull parameters and they all have a region in their parameter space in which massive gravitons are propagated
in a perturbatively unitary fashion around a maximally symmetric space-time. 

On AdS backgrounds there exist special points in the parameter space of these higher-derivative gravities at which (some of)
the linearized graviton modes coincide with each other. Those points are called critical points.
Typically, at such a critical point, some of
the massive gravitons degenerate with the massless gravitons and thus the spectrum no longer contains such massive gravitons.
Theories at critical points are referred to as critical gravities. Away from all the critical points,
the massless and massive graviton solutions show Brown--Henneaux boundary behavior \cite{Brown:1986nw} towards the AdS boundary. At
the critical point, the massive graviton solutions that have disappeared from the spectrum are replaced by
so-called logarithmic modes. The latter are characterized by a logarithmic boundary behavior that is more general than the
Brown--Henneaux one. 

The appearance of  logarithmic modes is important in formulating the AdS/CFT correspondence for critical gravities.
At the gravity side of the correspondence, one needs to specify boundary conditions for the excitations
that are kept in the gravity theory. The existence of linearized logarithmic modes indicates that, for critical gravities,
one can formulate consistent boundary conditions that
include excitations with logarithmic asymptotic behavior. The resulting dual CFT is conjectured to be a Logarithmic Conformal
Field Theory
(LCFT) \cite{Gurarie:1993xq,Flohr:2001zs,Kogan:2002mg}. The latter are characterized by the fact that their Hamiltonian
is non-diagonalizable and that their correlators contain logarithmic singularities. LCFTs contain operators that
have degenerate scaling dimensions with other operators that are referred to as logarithmic partners. Operators with degenerate
scaling dimensions organize themselves in Jordan cells, on which the Hamiltonian is non-diagonalizable. The dimension of the
Jordan cells is called the rank of the LCFT. LCFTs are typically
non-unitary, but have nevertheless been studied in condensed matter physics in a variety of contexts, such as
critical phenomena, percolation and turbulence. The conjecture that critical gravities with particular boundary conditions are
dual to LCFTs, was proposed in the context of critical TMG in \cite{Grumiller:2008qz}
and was later extended to critical three-dimensional NMG \cite{Liu:2009bk,Liu:2009kc}. More checks on the conjecture
were performed via explicit computation of two-point correlators
\cite{Skenderis:2009nt,Grumiller:2009mw,Grumiller:2009sn,Alishahiha:2010bw} and partition functions \cite{Gaberdiel:2010xv}.
A higher-dimensional analogue of critical NMG can be formulated \cite{Lu:2011zk,Deser:2011xc} and similar results on logarithmic
modes and their holographic consequences have been put forward in \cite{Alishahiha:2011yb,Bergshoeff:2011ri,Chen:2011in,Johansson:2012fs}.

An interesting question is whether one can formulate the AdS/CFT correspondence with a stricter set of boundary
conditions that do not allow all orders of logarithmic boundary behavior. On the CFT side, this could lead to a consistent
truncation of the LCFT. Apart from being interesting for the study of LCFTs, this  can also have implications
for the construction of toy models of quantum gravity; in particular when the truncated LCFT is unitary. Truncations
of critical gravities have been considered for critical TMG \cite{Li:2008dq} and (the higher-dimensional analogue of)
critical NMG \cite{Liu:2009bk,Liu:2009kc,Lu:2011zk}.
In both cases, the truncation amounts to imposing strict Brown--Henneaux boundary conditions \cite{Henneaux:1985tv}. In the case of
critical TMG, the truncation gives the so-called chiral gravity theory \cite{Li:2008dq}. This theory is dual to a chiral CFT,
implying that, at least classically, the theory admits a chiral, unitary sub-sector \cite{Maloney:2009ck}. In
the case of critical NMG and its higher-dimensional analogue, the truncated theory only describes a massless graviton with
zero on-shell energy.
Its black hole solutions also have zero energy and entropy. The theory thus seems trivial in the sense that
the truncation only retains the vacuum state, upon modding out zero energy states. It was suggested in \cite{Lu:2011ks}
that this feature is related to a recent proposal \cite{Maldacena:2011mk}, that states that four-dimensional
conformal gravity, with specific boundary conditions, is equivalent to Einstein gravity with a cosmological constant.

The truncations discussed above concern critical TMG and critical NMG, which are both dual to two-dimensional rank-2 LCFTs.
It was argued in \cite{Bergshoeff:2012sc}, how similar truncations can be defined for rank greater than two in the context
of a scalar field toy model. This scalar field toy model describes $r$ coupled scalar fields with degenerate masses and
corresponds to a critical point of a theory with $r$ scalars with non-degenerate masses. In particular, at the critical point,
the toy model not only contains a massive scalar solution, but also $r-1$ solutions that exhibit logarithmic boundary behavior.
For every power $n = 1, \cdots, r-1$, there is one solution that falls off as $\log^n$. Such solutions are referred to as
`$\log^n$ modes'. For boundary conditions that keep all $\log^n$ modes, the two-point functions of the dual CFT were shown
to correspond to those of a rank-$r$ LCFT. This model is a toy model for a parity even theory and it was argued that in
this case\footnote{The parity
odd case is slightly different in the sense that the left-moving and right-moving sectors can behave differently.
For instance, in the case of critical TMG one sector corresponds to an ordinary CFT, while the other sector
corresponds to an LCFT of rank 2. One can then apply the truncation to the LCFT sector alone. The full resulting
theory is still non-trivial due to the presence of the ordinary CFT sector.} there is a qualitative difference
between the cases of even and odd rank, when considering truncations of the dual LCFT. For even ranks, one can define a
truncation such that the resulting theory has trivial two-point correlators and only seems to contain null
states. This is analogous to what happens for (the higher-dimensional analogue of) critical NMG. For odd ranks, a similarly
defined truncation leads to a theory that has one two-point
function whose structure is that of an ordinary CFT. In addition to that, the theory also contains null states.
This indicates that odd rank LCFTs might allow for a non-trivial truncation.

The results mentioned in the previous paragraph were obtained in the context of a spin-0 toy model. Although
interesting in its own right, such a toy model is limited in some respects. Most notably, the model is non-interacting
and there is no organizational principle, such as gauge invariance, that can suggest interesting
interactions. In order to study the truncation procedure in the presence of interactions, one needs to go beyond
this scalar field toy model. It is thus interesting to look at a three-dimensional gravity realization of
two-dimensional, odd rank LCFTs. 

The precise form of such a spin-2 realization depends on whether one considers parity even or odd models. Since
the number of linearized solutions propagated by the higher-derivative theory is essentially given by the order
in derivatives of the theory, one can already construct a rank-3, parity odd theory in the context of
four-derivative gravity, i.e.~in the context of GMG. Indeed there exists a critical point in the GMG parameter
space where the theory propagates one left-moving massless boundary graviton, as well as a right-moving massless boundary graviton
and two associated logarithmic modes, with $\log$ and $\log^2$ boundary behavior respectively \cite{Liu:2009pha}.
This gives a total of four modes\footnote{For parity odd theories we count all helicity states separately, while for parity
even models we will refer to the two helicity states as one mode.}, as expected of a four-derivative theory.
This critical point is sometimes called
`tricritical' and this critical version of GMG is correspondingly called `tricritical GMG'. It was shown in
\cite{Grumiller:2010tj,Bertin:2011jk} that the structure of the dual CFT is consistent with that of a parity
violating LCFT of rank 3. Note that the parity oddness of tricritical GMG is reflected in the fact that the
logarithmic modes are only associated to the left-moving massless graviton. In a parity even theory, both
left- and right-moving massless gravitons need to degenerate with the same number of $\log^n$ modes. One can
thus see that in order to get a parity even, critical theory that propagates $\log^n$ modes with $n \geq2$,
one needs to consider theories with more than four derivatives. In particular, to get a tricritical model,
that propagates massless boundary gravitons, $\log$ and $\log^2$ modes, one has to look at six-derivative gravity models.
This was already suggested in \cite{Bergshoeff:2012sc}, where also the expected form of the linearized equations
of motion at the tricritical point was given. Higher-derivative gravities in $d \geq 4$, that have critical points
that can be conjectured to be dual to higher-rank LCFTs in more than two dimensions, were considered in \cite{Nutma:2012ss}.

In this paper, we look at a specific parity even, tricritical, six-derivative gravity model, that we
call Parity Even Tricritical (PET) gravity. As suggested in \cite{Bergshoeff:2012sc}, we start
from a three-dimensional gravity theory that contains generic $R^2$ and $R \square R$ terms, where $R$ denotes
the Ricci scalar or tensor. We linearize around an $\mathrm{AdS}_3$ background, and we show that for a certain choice of
the parameters, one obtains a fully non-linear gravity theory with a tricritical point,
at which massless boundary gravitons, $\log$ and $\log^2$ modes are propagated (at the linearized level). The existence and properties of these
logarithmic modes lead one to conjecture that the CFT-dual of PET gravity is an LCFT of rank 3, if appropriate boundary conditions
that include excitations with $\log^2$ boundary behavior are adopted. The structure
of the two-point correlators of such an LCFT is of the form as obtained in \cite{Bergshoeff:2012sc} in the
context of the scalar field toy model and similar remarks about truncating the odd rank LCFT by restricting
the boundary conditions can thus be made. Here we rephrase this truncation procedure in a different
manner, analogously to what was done in \cite{Maloney:2009ck} in the context of critical TMG. We 
consider the conserved charges associated to (asymptotic) symmetries and we show that the truncation procedure
is equivalent to restricting to a zero charge sub-sector of the theory. This formulation is useful in the discussion of the
consistency of the truncation. Indeed, the introduction of interactions
can spoil the consistency of the truncation, as the restricted boundary conditions are not necessarily preserved
under time evolution. Rephrasing the truncation procedure as a restriction to a zero
charge sub-sector allows one to use charge conservation arguments to guarantee the consistency of the truncation at the classical level. 

The outline of the paper is as follows. In section \ref{PETsection}, the PET model will be introduced as the
tricritical version of a parity even, six-derivative gravity theory in three dimensions. The linearization of
this theory will be given, along with a different formulation that is second order in derivatives but
involves two auxiliary fields.  Section \ref{BHsection} contains a discussion on black hole type
solutions of non-linear PET gravity. In section \ref{PETCFTsection}, we consider solutions of the linearized
equations of motion. We show that the PET model exhibits massless graviton solutions,
along with $\log$ and $\log^2$ solutions. We give arguments that support the conjecture that PET gravity,
with boundary conditions that include excitations with asymptotic $\log^2$ behavior, is dual to a rank-3 LCFT and we comment on
the structure of the dual LCFT. The boundary stress tensor, the central charges and the new anomaly of
the dual LCFT are calculated on the gravity side, and the structure of the two-point correlators will be given.
In section \ref{truncsection}, we consider the truncation of \cite{Bergshoeff:2012sc} in the PET model.
We calculate the conserved charges associated to (asymptotic) symmetries and we show that the
truncation can be rephrased as a restriction to a zero charge sub-sector. We comment on the form of the
two-point correlators in the truncated theory. We conclude in section \ref{conclsection} with a discussion of the obtained results. As mentioned above, GMG also exhibits
a tricritical point, where the theory is conjectured to be dual to a rank-3 LCFT. In tricritical GMG, a similar
truncation can be made, and again this truncation can be rephrased as a restriction to a zero charge sub-sector.
The results necessary to discuss this truncation in this parity odd example, have been given in \cite{Liu:2009pha}.
In appendix \ref{GMGapp} we summarize these results to illustrate the truncation procedure in a parity odd setting.
Appendix \ref{stresstensorapp}  contains technical details on the calculation of the boundary stress tensor of PET gravity. In appendix \ref{Energyapp} we calculate the on-shell energy of the linearized modes in the theory.

\section{A Parity Even Tricritical (PET) model} \label{PETsection}

As outlined in the introduction, in this section we consider a three-dimensional gravity theory with generic $R^2$
and $R \square R$ terms (with $R$ either the Ricci tensor or scalar). We linearize this theory around an $\mathrm{AdS}_3$
background and we restrict the parameter space such that the theory propagates only two massive spin-2 excitations, in
addition to a massless boundary graviton. We will show that there is a tricritical point in the restricted parameter space, where
both massive excitations become massless and degenerate with the massless mode. The PET model is then defined as the
gravity theory at this tricritical point. The PET model is of sixth order in derivatives. For some applications,
it is useful to have a formulation that is of second order in derivatives. This can be done at the expense of introducing
auxiliary fields. The auxiliary field formulation of our model will be given in subsection \ref{auxfieldsssec}. Finally, in section
\ref{BHsection}, we will discuss black hole type solutions of PET gravity.

\subsection{A six-derivative gravity model and its tricritical point}\label{PETmodelsection}

In three dimensions, the most general Einstein--Hilbert action supplemented with a cosmological parameter, $R^2$
and $R\square R$ terms,\footnote{Note that this is not the most general six-derivative action. Generic terms that
involve cubic powers of the curvatures could be added. Such terms would lead to the introduction of extra dimensionfull 
parameters in the model. For simplicity, we will only consider the case where the six-derivative terms are of the
form $R \square R$ in this paper.} is
\begin{equation}
S = \frac{1}{16 \pi G} \int \extd^3 x \sqrt{-g} \left\{\sigma R - 2\Lambda_0+ \alpha R^2 + \beta R_{\mu\nu}R^{\mu\nu}  + \cL_{R\square R} \right\} \,,
\label{actionsix}
\end{equation}
where
\begin{align}  
\cL_{R\square R} & = b_1 \nabla_{\mu}R \nabla^{\mu} R + b_2 \nabla_{\rho} R_{\mu\nu} \nabla^{\rho} R^{\mu\nu} \,.
\end{align}
The dimensionless parameter $\sigma$ is given by  $0, \pm 1$, whereas $\Lambda_0$ is a cosmological parameter. The parameters $\alpha,\beta$ are arbitrary
parameters of dimension inverse mass squared and $b_1,b_2$ are arbitrary parameters with dimensions of inverse
mass to the fourth. The theory has sixth order equations of motion that read
\begin{equation} \label{eompet}
\sigma G_{\mu\nu} + \Lambda_0 g_{\mu\nu} + E_{\mu\nu} + H_{\mu\nu} = 0 \,,
\end{equation}
with
\begin{align}
E_{\mu\nu}  =& \, \alpha \left(2 RR_{\mu\nu} - \frac12 g_{\mu\nu}R^2 + 2g_{\mu\nu} \square R - 2 \nabla_{\mu} \nabla_{\nu} R \right) 
 + \beta \bigg(   \frac32 g_{\mu\nu} R_{\rho\sigma} R^{\rho\sigma} \\ \nonumber
 & -4R_{\mu}^{\rho} R_{\nu\rho} + \square R_{\mu\nu} + \frac12 g_{\mu\nu}\square R - \nabla_{\mu}\nabla_{\nu}R + 3 RR_{\mu\nu} - g_{\mu\nu}R^2 \bigg) \nonumber \,,
 \\
H_{\mu\nu} = & \, b_1 \left( \nabla_{\mu}R\nabla_{\nu}R  - 2 R_{\mu\nu}\square R - \frac12 g_{\mu\nu} \nabla_{\alpha}R \nabla^{\alpha}R - 2(g_{\mu\nu}\square^2 - \nabla_{\mu}\nabla_{\nu}\square) R \right) \\ \nonumber
& + b_2 \bigg( \nabla_{\mu}R_{\rho\sigma}\nabla_{\nu} R^{\rho\sigma} - \frac12 g_{\mu\nu} \nabla_{\alpha} R_{\rho\sigma} \nabla^{\alpha} R^{\rho\sigma} - \square^2R_{\mu\nu} - g_{\mu\nu}\nabla^{\rho}\nabla^{\sigma}\square R_{\rho\sigma} \\ \nonumber
& + 2 \nabla^{\rho}\nabla_{(\mu}\square R_{\nu)\rho} + 2 \nabla^{\rho}R_{\rho\sigma}\nabla_{(\mu}R_{\nu)}^{\sigma} + 2 R_{\rho\sigma}\nabla^{\rho}\nabla_{(\mu}R_{\nu)}^{\sigma} - 2 R_{\sigma(\mu} \square R_{\nu)}^{\sigma} \\ \nonumber
& - 2 \nabla_{\rho} R_{\sigma(\mu}\nabla_{\nu)}R^{\rho\sigma} - 2 R_{\sigma (\mu}\nabla^{\rho}\nabla_{\nu)} R_{\rho}^{\sigma} \bigg) \,.
\end{align}
The equations of motion \eqref{eompet} allow for $\mathrm{AdS}_3$ solutions with cosmological constant $\Lambda$
that obeys the equation
\begin{equation}
\sigma \Lambda -  \Lambda_0   - 6 \Lambda^2 \alpha - 2 \Lambda^2 \beta = 0 \,.
\end{equation} 
We will now consider the linearization of the equations of motion \eqref{eompet} around such a background.
Denoting background quantities with a bar,  the metric can be expanded around its background $\mathrm{AdS}_3$
value $\bar{g}_{\mu \nu}$ as
\begin{equation}
g_{\mu \nu} = \bar{g}_{\mu \nu} + h_{\mu \nu} \,.
\end{equation}
The background curvature quantities are
\begin{equation} 
\R_{\mu\nu\rho\sigma}  = 2\Lambda \bg_{\mu[\rho}\bg_{\sigma]\nu}\,, \quad 
\R_{\mu\nu}  = 2\Lambda \bg_{\mu\nu}\,, \quad \R  = 6\Lambda \,, \quad \bar{G}_{\mu\nu} = - \Lambda \bg_{\mu\nu}\,.
\end{equation}
The linearized equations of motion for the metric fluctuation $h_{\mu \nu}$ are then given by
\begin{align} \label{lineom}
0 = & \, \bar{\sigma} \cG_{\mu\nu} - (2\beta  - 4 \Lambda b_2) \cG_{\mu\nu}(\cG(h)) - 4 b_2 \cG_{\mu\nu}(\cG(\cG(h)))
\\ \nonumber
& + (2\alpha + \frac12 \beta ) \left(\bg_{\mu\nu} \bBox - \bnabla_{\mu} \bnabla_{\nu} + 2\Lambda\bg_{\mu\nu}\right) R^{(1)}  \\ \nonumber
& - (2b_1 + b_2) \left( \bg_{\mu\nu} \bBox - \bnabla_{\mu} \bnabla_{\nu} + 2\Lambda\bg_{\mu\nu} \right) \bBox R^{(1)}\,,
\end{align}
where the constant $\bar{\sigma}$ is given by
\begin{equation}
\bar{\sigma} = \sigma + 12\Lambda\alpha + 4 \Lambda \beta \,.
\end{equation}
The linearized Einstein operator $\cG_{\mu \nu}$ is expressed in terms of the linearized Ricci tensor $R_{\mu\nu}^{(1)}$
and linearized Ricci scalar $R^{(1)}$
\begin{equation} \label{linRicci}
R_{\mu\nu}^{(1)} =  \bnabla^{\rho}\bnabla_{(\mu} h_{\nu) \rho} - \frac12 \bBox h_{\mu\nu} - \frac12 \bnabla_{\mu} \bnabla_{\nu} h\,, \quad
R^{(1)} =  - \bBox h + \bnabla^{\rho} \bnabla^{\sigma} h_{\rho\sigma} - 2\Lambda  h\,,
\end{equation}
as follows:
\begin{equation} \label{lineinstein}
\cG_{\mu\nu} =  R^{(1)}_{\mu\nu} - \frac12 \bg_{\mu\nu} R^{(1)} - 2\Lambda h_{\mu\nu}\,.
\end{equation}
Note that $\cG_{\mu \nu}$ is invariant under linearized general coordinate transformations and that it obeys
\begin{equation}
\bnabla^\mu \cG_{\mu \nu} = 0 \,.
\end{equation}
The trace of the linearized equations of motion \eqref{lineom} is given by
\begin{align}
	\label{lintraceeqm}
0=& \,\left(-\frac12 \sigma + 6\Lambda \alpha + 2 \Lambda \beta \right) R^{(1)} + \left( 4 \alpha + \frac32 \beta - 12 \Lambda b_1 - 5 \Lambda b_2 \right) \bBox R^{(1)} \\ \nonumber
& - \left(4b_1 + \frac32 b_2 \right) \bBox^2 R^{(1)}\,.
\end{align}
In order to avoid propagating scalar degrees of freedom, we will restrict our attention to parameters that satisfy
\begin{equation} \label{firstrestr}
b_1 = -\frac38 b_2 \,, \qquad \alpha = \frac{\Lambda}{8} b_2 - \frac38 \beta \,,
\end{equation}
and we will moreover assume that 
\begin{equation} \label{secondrestr}
-\frac{\sigma}{2} + \frac34 \Lambda^2 b_2 - \frac{\Lambda}{4} \beta \neq 0 \,.
\end{equation}
The first conditions \eqref{firstrestr} then ensure that \eqref{lintraceeqm} does not contain any $\bBox R^{(1)}$ and $\bBox^2 R^{(1)}$ terms, while the assumption \eqref{secondrestr} entails that \eqref{lintraceeqm} implies that $R^{(1)} = 0$. If we subsequently choose the gauge
\begin{equation} \label{gaugechoice}
\bar{\nabla}^\mu h_{\mu \nu} = \bar{\nabla}_\nu h \,,
\end{equation}
we find that $R^{(1)}$ simplifies to $R^{(1)} = - 2 \Lambda h$ and thus $h=0$.
Hence the metric perturbations in the gauge \eqref{gaugechoice} are transverse-traceless:
\begin{equation} \label{ttgauge}
\bar{\nabla}^\mu h_{\mu \nu} = h = 0 \,.
\end{equation}
The linearized equations of motion \eqref{lineom} then simplify to
\begin{eqnarray} \label{lineom2}
\bar{\sigma} \cG_{\mu\nu} - (2\beta - 4 \Lambda b_2) \cG_{\mu\nu}(\cG(h)) - 4 b_2 \cG_{\mu\nu}(\cG(\cG(h))) = 0 \,,
\end{eqnarray}
where the gauge-fixed linearized Einstein operator is given by
\begin{equation}
\cG_{\mu\nu} = - \frac12 \left(  \bBox - 2\Lambda \right)h_{\mu\nu}\,.
\label{gauge-fiexedLinearEinstein}
\end{equation}
The linearized equations of motion \eqref{lineom2} can be rewritten as \footnote{These linearized e.o.m.~are contained within the
class of theories considered in \cite{Nutma:2012ss}.}               
\begin{align}\label{lineomgauge}
\left(\bBox - 2\Lambda \right) \left(\bBox - 2 \Lambda - M_+^2\right)\left(\bBox - 2 \Lambda - M_-^2\right)h_{\mu\nu} = 0 \,,
\end{align}
where the mass parameters $M_{\pm}$ are given by
\begin{equation}\label{Masses}
M_{\pm}^2 = \frac{\beta}{2b_2} - \Lambda  \pm \frac{1}{2b_2}\sqrt{10 b_2^2 \Lambda ^2-6 b_2 \beta  \Lambda +4 b_2 \sigma +\beta^2} \,.
\end{equation}
From equation \eqref{lineomgauge} it is clear that our class of theories, with the restrictions \eqref{firstrestr} on the
parameters, has solutions that correspond to a massless spin-2 mode $h^{(0)}_{\mu \nu}$ and two massive spin-2 modes
$h^{(M_\pm)}_{\mu \nu}$ that satisfy the following Klein--Gordon-type equations:
\begin{equation}
\left(\bBox - 2\Lambda \right)h^{(0)}_{\mu\nu} = 0\,, \qquad \left(\bBox - 2 \Lambda - M_{\pm}^2\right)h^{(M_{\pm})}_{\mu\nu} =0\,.
\end{equation}
The case for which 
\begin{equation} \label{tricritpoint}
\beta = - 4\frac{\sigma}{\Lambda} \quad {\rm and} \quad b_2 = - 2 \frac{\sigma}{\Lambda^2}
\end{equation}
is special. At this point in parameter space $\bar{\sigma} = 0$ and $M_{\pm}^2 = 0$. This point corresponds to a
critical point in parameter space, where both massive modes degenerate with the massless mode. Since
this degeneracy is threefold, the point \eqref{tricritpoint} corresponds to a tricritical point. The linearized
equations of motion at this tricritical point assume the simple form\footnote{Note that in \cite{Lu:2011ks} a six-derivative
theory with similar e.o.m.~has been considered.}
\begin{equation} \label{eomcrit}
\mathcal{G}_{\mu \nu}(\mathcal{G}(\mathcal{G}(h))) = 0 \,.
\end{equation}
This corresponds to the spin-2 version of the equations of motion of the rank-3 scalar field model, discussed in
\cite{Bergshoeff:2012sc}. The theory at this tricritical point will be called Parity Even Tricritical gravity (PET gravity). 

\begin{figure}
\centering
\includegraphics[width=0.8\textwidth]{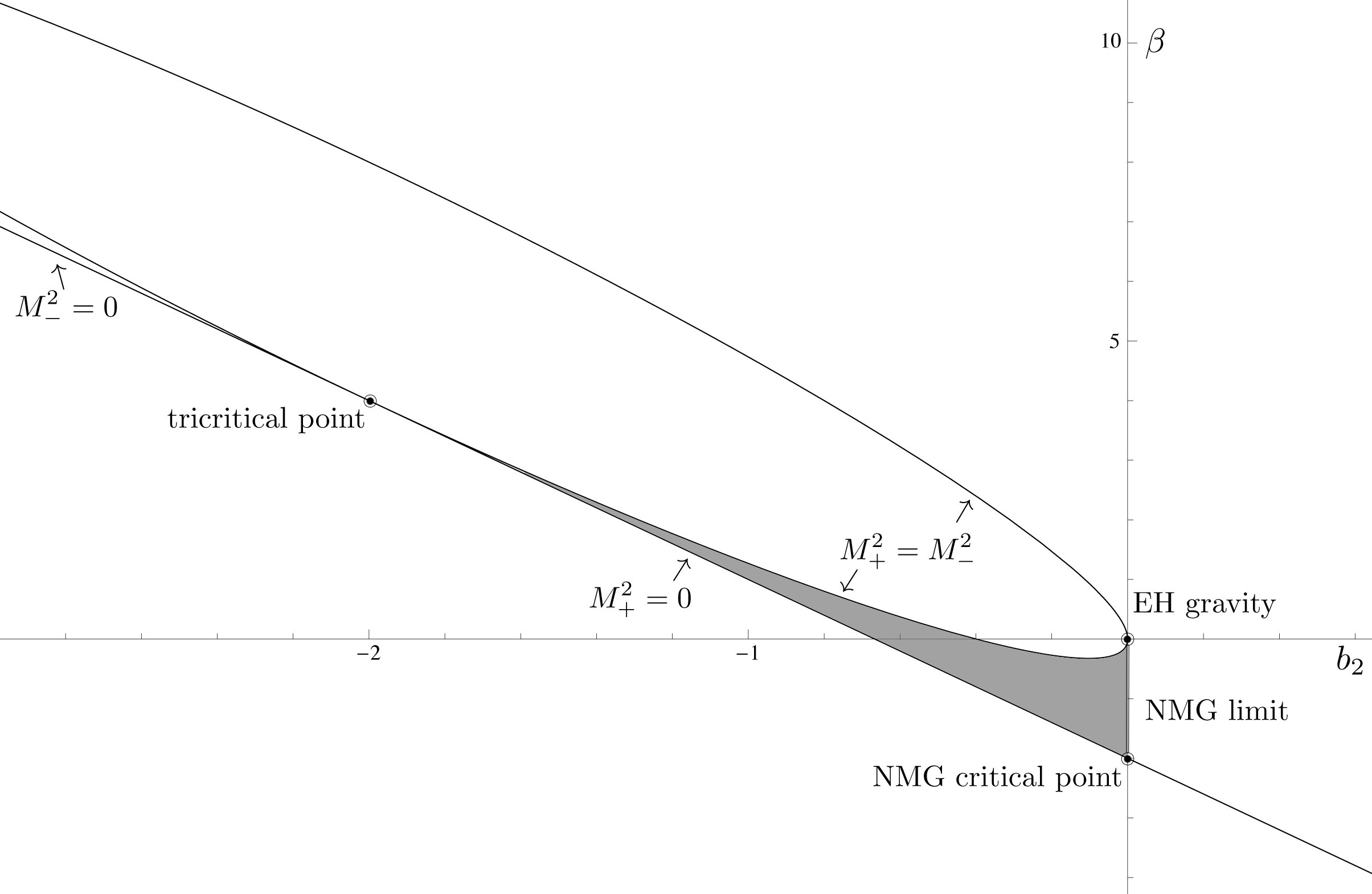}
\caption{\small The parameter space of the sixth order gravity model with $\Lambda = -1$ and $\sigma=1$. A similar figure
can be made for $\sigma=-1$, where the figure is mirrored along the $\beta$ and $b_2$-axis and $M_+$ and $M_-$ are interchanged.
The shaded region denotes where
the mass squared of both massive modes is either zero of positive. The limit $b_2 \to 0$ is the NMG limit where
one of the masses becomes infinite and the other takes values between zero (NMG critical point) and infinity (Einstein gravity limit).}
\label{parameterspace}
\end{figure}

Apart from this tricritical point, there are many other critical points in the $(\beta,b_2)$ parameter space of
the presented six-derivative model, where degeneracies take place. In particular, there is a critical curve, defined via
\begin{equation}
	\label{critcurve}
10 b_2^2 \Lambda ^2-6 b_2 \beta  \Lambda +4 b_2 \sigma +\beta^2 = 0 \,,
\end{equation}
where both massive gravitons degenerate with each other, i.e.~$M_+^2 = M_-^2$. Similarly, there is a critical line, defined via
\begin{equation} \label{critline}
3 \Lambda^2 b_2 - \beta \Lambda + 2 \sigma = 0 \,,
\end{equation}
where only one of the massive gravitons becomes massless (e.g.~where $M_+^2 = 0$, while generically $M_-^2 \neq 0$ or vice versa). The
situation is summarized in figure \ref{parameterspace} where the parameter space for the sixth order gravity model is
displayed. The requirement that both the masses are real valued ($M_{\pm}^2 \geq 0$) implies that the parameters $\beta$
and $b_2$ may only take on values within the shaded region. The borders of the shaded region are the critical lines
\eqref{critcurve} and \eqref{critline}, and the $\beta$-axis.
The $b_2\to0$ limit corresponds to NMG, where one of the masses becomes infinite while the other stays finite and
corresponds to the massive graviton of NMG. The corners of the triangle denote three special limits of the theory. 
The origin is just cosmological Einstein--Hilbert gravity, where both masses become infinite and hence both massive gravitons decouple.
The other point on the $\beta$-axis is the NMG critical point.
Here one of the masses decouples and the other becomes zero. The $\beta$ parameter now takes on the NMG critical value
$\beta = 1/m^2 = 2\sigma/ \Lambda$. The third corner at $\beta = -4 \sigma/ \Lambda$ and $b_2 = -2 \sigma / \Lambda^2$ is
the tricritical point, discussed above.

\subsection{Auxiliary field formulation} \label{auxfieldsssec}

The above PET model is of sixth order in derivatives. For some purposes, such as e.g.~the calculation of the
boundary stress tensor, it is easier to work with a two-derivative action. It is possible to reformulate the
action \eqref{actionsix}, subject to the parameter choice \eqref{firstrestr}, as a two-derivative theory upon
the introduction of two auxiliary fields $f_{\mu\nu}$ and $\lambda_{\mu\nu}$. The action in terms of the metric
and the two auxiliary fields is given by
\begin{align}\begin{split}\label{aux2}
	S = & \frac{1}{16 \pi G} \int \rmd^3 x \sqrt{-g} \, \bigg\{ \sigma R - 2\Lambda_0 + f^{\mu\nu} G_{\mu\nu} 
	- \left( \lambda^{\mu\nu} f{}_{\mu\nu} - \lambda f\right) 
	+ \beta (\lambda^{\mu\nu}\lambda_{\mu\nu} - \lambda^2) \\
	& + 2\Lambda b_2 \lambda^2 -b_2\left( \lambda^{\mu\nu} \square \lambda_{\mu\nu}
        - \lambda \square \lambda \right) + 2b_2 \left( \lambda^{\mu\nu}\nabla_{(\mu}\nabla^{\rho}\lambda_{\nu)\rho}
        - \lambda \nabla^{\mu}\nabla^{\nu}\lambda_{\mu\nu} \right) \bigg\} \,,
\end{split}\end{align}
where $\lambda=\lambda_{\mu\nu}g^{\mu\nu}$ and $f=f_{\mu\nu}g^{\mu\nu}$ are the traces of the resp.~auxiliary fields.
The equations of motion for the auxiliary fields are
\begin{align}
 \lambda_{\mu\nu} =&\,  R_{\mu\nu} - \frac14 g_{\mu\nu} R\,, \\
 f{}_{\mu\nu} =&\, 2\beta  \lambda_{\mu\nu} - 2 \Lambda b_2 g_{\mu\nu} \lambda  
  + 2b_2 \Big( 2 \nabla_{(\mu}\nabla^{\rho}\lambda_{\nu)\rho} -\Box \lambda_{\mu\nu}
  - \frac12 g_{\mu\nu} \nabla^{\rho}\nabla^{\sigma}\lambda_{\rho\sigma} \\ 
 & \nonumber - \nabla_{\mu}\nabla_{\nu} \lambda + \frac12 g_{\mu\nu}\square \lambda \Big)\,.
\end{align}
Substituting these expressions into (\ref{aux2}) gives the action \eqref{actionsix} with the parameters $\alpha$
and $b_1$ given by \eqref{firstrestr}, so the two actions \eqref{actionsix} and \eqref{aux2} are indeed classically
equivalent, in the parameter range of interest.

We proceed to linearize this action around an AdS background, where we take the following linearization ansatz:
\begin{align}
g_{\mu\nu} & = \bg_{\mu\nu} +  h_{\mu\nu} \,, \\
\lambda_{\mu\nu} & = \frac{\Lambda}{2}\left(\bg_{\mu\nu} +  h_{\mu\nu} \right) + k_1{}_{\mu\nu} \,, \\
f_{\mu\nu} & = \left(\Lambda\beta - 3\Lambda^2 b_2 \right) \left(\bg_{\mu\nu} +  h_{\mu\nu} \right) + k_2{}_{\mu\nu} \,.
\end{align}
Plugging this into the action \eqref{aux2} and keeping the terms that are quadratic in the fields $h_{\mu \nu}$, $k_1{}_{\mu\nu}$ and $k_2{}_{\mu\nu}$, we find the following linearized action:
\begin{align} \label{linShk}
\cL^{(2)} = &\, - \frac12 \bar{\sigma} h^{\mu\nu} \cG_{\mu\nu}(h) + k^{\mu\nu}_2 \cG_{\mu\nu}(h) + 2b_2 k_1^{\mu\nu} \cG_{\mu\nu}(k_1) \\ 
& - (2\Lambda b_2 -\beta) \left(k_1^{\mu\nu} k_{1\mu\nu} - k_1^2\right) - \left( k_1^{\mu\nu}k_{2\mu\nu} - k_1 k_2 \right)\,.
\nonumber & 
\end{align}
Assuming that $\bar{\sigma} \neq 0$, the linearized Lagrangian may be diagonalized. After the field redefinition\footnote{One may
also redefine the fields with $M_-$ replaced by $M_+$. This will also lead to a diagonalized Lagrangian, but the roles of $k_1$
and $k_2$ will be interchanged in \eqref{L2diag}.}
\begin{align}\label{shift1}
h_{\mu\nu} = & \; h'_{\mu\nu} + \frac{2 b_2 M_-^2}{\bar{\sigma}} k'_1{}_{\mu\nu} + \frac{1}{\bar{\sigma}} k'_2{}_{\mu\nu}\,,  \\
k_1{}_{\mu\nu} = & \; k'_1{}_{\mu\nu} - \frac{M_-^2}{2\bar{\sigma}}k'_2{}_{\mu\nu}\,, \label{shift2}\\
k_2{}_{\mu\nu} = & \; k'_2{}_{\mu\nu} + 2b_2 M_-^2 k'_1{}_{\mu\nu}\,, \label{shift3}
\end{align}
equation \eqref{linShk} becomes a Lagrangian for a massless spin-2 field $h'_{\mu\nu}$ and two massive spin-2 fields with mass $M_{\pm}^2$:
\begin{align}\label{L2diag}
\cL^{(2)} = &\, - \frac12 \bar{\sigma} h'{}^{\mu\nu} \cG_{\mu\nu}(h') \nonumber \\
& + \frac{4 b_2}{\bar{\sigma}}\left(\bar{\sigma} + b_2M_-^4  \right) \left[  \frac12 k'_1{}^{\mu\nu} \cG_{\mu\nu}(k'_1) - \frac14 M_+^2  \left(k'_1{}^{\mu\nu} k'_{1\mu\nu} - k'_1{}^2\right) \right]  \\ 
& + \frac{1}{\bar{\sigma}^2}\left( \bar{\sigma} + b_2M_-^4 \right) \left[  \frac12 k'_2{}^{\mu\nu} \cG_{\mu\nu}(k'_2) - \frac14 M_-^2  \left(k'_2{}^{\mu\nu} k'_{2\mu\nu} - k'_2{}^2\right) \right]  \,.
\nonumber 
\end{align}
In order to make sure that there are no ghosts, we must demand that the kinetic terms in \eqref{L2diag} all have the same sign. One can
see that for $\bar{\sigma} \neq 0$ the absence of ghosts can not be reconciled with the reality of $M_{\pm}^2$. Away from the critical lines,
at least one of the modes is a ghost. The same result may be derived from the expression of the on-shell energy of the massless
and massive modes given in appendix \ref{Energyapp}. The appearance of ghosts away from the critical lines is consistent with results
found for higher-dimensional and higher-rank critical gravity theories in \cite{Nutma:2012ss}.

At the critical line and the tricritical point, the field redefinitions leading to this Lagrangian are not well-defined and the
Lagrangian \eqref{linShk} is non-diagonalizable. Let us first consider the critical line \eqref{critline}. Here one of the massive
modes degenerates with the massless mode and one expects that \eqref{linShk} may be written as a Fierz--Pauli Lagrangian for the
remaining massive mode plus a part which resembles the linearized Lagrangian of critical NMG. Indeed, after a field redefinition
\begin{align}\label{shift4}
h_{\mu\nu} = & \; h''_{\mu\nu} - 4 b_2 \alpha k''_1{}_{\mu\nu}- 2 b_2 \alpha^2 k''_2{}_{\mu\nu}\,,  \\
k_1{}_{\mu\nu} = & \; k''_1{}_{\mu\nu}+ \alpha k''_2{}_{\mu\nu}\,, \\
k_2{}_{\mu\nu} = & \; k''_2{}_{\mu\nu}\,, 
\end{align}
with $\alpha = \frac12 (\frac{2\sigma}{\Lambda} + \Lambda b_2)^{-1}$, the Lagrangian \eqref{linShk} reduces to
\begin{align}\label{L2critline}
\cL^{(2)} = &\,  k''_2{}^{\mu\nu} \cG_{\mu\nu}(h'')  - \frac14 \left(\frac{2\sigma}{\Lambda} +\Lambda b_2 \right)^{-1} \left( k''_2{}^{\mu\nu}k''_{2\mu\nu} - k''_2{}^2 \right) \\
\nonumber &  + 4b_2 \left[ \frac12 k''_1{}^{\mu\nu} \cG_{\mu\nu}(k''_1) - \frac14 M'^2 \left(k''_1{}^{\mu\nu} k''_{1\mu\nu} - k''_1{}^2\right) \right]\,,
\nonumber & 
\end{align}
with $M'^2 = - (\frac{2\sigma}{b_2 \Lambda} + \Lambda )$. At the tricritical point \eqref{tricritpoint} this semi-diagonalization procedure breaks down and we must work with the non-diagonal action:
\begin{equation} \label{L2tricrit}
\cL^{(2)} = \, k^{\mu\nu}_2 \cG_{\mu\nu}(h) + 2b_2 k_1^{\mu\nu} \cG_{\mu\nu}(k_1)  - \left( k_1^{\mu\nu}k_{2\mu\nu} - k_1 k_2 \right)\,.
 \end{equation}
Let us now show how this linearized action leads to the linearized equations of motion of \eqref{eomcrit}. The equations of motion derived from \eqref{L2tricrit} are:
\begin{align} 
  \cG_{\mu\nu}(k_2) & = 0 \label{w4} \,, \\
  4b_2 \cG_{\mu\nu} (k_1) & =  \left( k_{2\mu\nu} - k_2 \bg_{\mu\nu} \right)  \label{w5} \,, \\
\cG_{\mu\nu}(h) & = \left( k_{1\mu\nu} - k_1 \bg_{\mu\nu} \right) \,. \label{w6}
\end{align}
Since $\nabla^{\mu} \cG_{\mu\nu}(k_1) = 0$, \eqref{w5} implies
\begin{equation}
\nabla^{\mu} k_{2\mu\nu} = \nabla_{\nu} k_2 \,.
\end{equation}
Together with the trace of \eqref{w4} we can conclude that $k_2 =0$ and thus
\begin{equation}
\cG_{\mu\nu}(\cG(k_1)) = 0 \label{w7} \,.
\end{equation}
Also, $\nabla^{\mu} \cG_{\mu\nu}(h) = 0$, so $\nabla^{\mu} k_{1\mu\nu} = \nabla_{\nu} k_1$ which, together
with \eqref{w7}, implies that $\frac12 \bBox k_1 + \Lambda k_1 = 0$. Using these identities we may rewrite
the equations of motion as
\begin{equation}
\cG_{\mu\nu}(\cG(\cG(h))) = 0\,,
\end{equation}
which is  what we obtained before in \eqref{eomcrit}.

\subsection{Non-linear solutions of PET gravity} \label{BHsection}

In this section, we will have a look at some solutions of the full non-linear theory that have log and log$^2$
asymptotics and can be related to black hole type solutions. In particular, we will first look at the BTZ black hole
\cite{Banados:1992wn}. The metric for the rotating BTZ black hole is given by
\begin{align}\begin{split}
 ds^2=\frac{dr^2}{N^2(r)}-N^2(r) dt^2+r^2(N_\phi(r) dt-d\phi)^2 \,, \\
 N^2(r)=\frac{r^2}{\ell^2}-8G{\rm m}+\frac{16G^2{\rm m}^2\ell^2}{r^2} \,, \qquad\qquad N_\phi(r)=\frac{4G{\rm j}}{r^2 } \,,
\end{split}\end{align}
where m and j are constants. This is a solution of the full sixth order theory for any m and j. The asymptotic form
of the BTZ black hole in Fefferman--Graham coordinates as an expansion around the conformal boundary $y=0$, is given
by \cite{Kraus:2005zm}
\begin{align}\label{BTZmetric}
 ds^2 = \frac{\ell^2dy^2}{y^2} -\frac{1}{y^2}\big(dt^2-\ell^2d\phi^2\big)
            +4G\big({\rm m}\, dt^2+{\rm m}\,\ell^2d\phi^2-2\,{\rm j}\,dt\, d\phi\big)  +\mathcal{O}(y^2)\,.
\end{align}
Here and in the following, we use the AdS length
$\ell = 1/\sqrt{-\Lambda}$. The mass and angular momentum of this BTZ black hole can be calculated using the boundary stress
tensor. The calculation of the boundary stress tensor will be given in appendix \ref{stresstensorapp}, while the
result will be discussed in section \ref{stresstenssec}. Anticipating that discussion, here we give the results
for the mass and angular momentum obtained from the boundary stress tensor\footnote{The masses and angular momenta
are given by boundary integrals of components of the stress tensor. Since none
of the components depends on the boundary coordinates they are simply given by
\begin{align}\label{bhMJ}
 M=2\pi\ell\,T_{tt} \qquad {\rm and} \qquad J=-2\pi\ell\,T_{t\phi} \,.
\end{align}}
for the rotating BTZ black hole:
\begin{align}
 M_{\rm BTZ}=\frac{\rm m}{2}\Big(2\sigma+\frac{\beta}{\ell^2}+\frac{3b_2}{\ell^4}\Big) \quad{\rm and}\quad
 J_{\rm BTZ}=\frac{\rm j}{2}\Big(2\sigma+\frac{\beta}{\ell^2}+\frac{3b_2}{\ell^4}\Big) \,.
\end{align}
Note that for the extremal case, when $\ell M_{\rm BTZ} = - J_{\rm BTZ}$, we also have that the constants $\rm m$ and
$\rm j$ obey $\rm j = - \ell \rm m$. In that case and furthermore restricting to the critical points and lines specified
above, the leading order terms of the metric \eqref{BTZmetric} can be dressed up with logarithmic asymptotics \cite{Clement:2009ka}, namely one can
find solutions of the form
\begin{equation}\label{logbhmetric}
 ds^2 = \frac{\ell^2dy^2}{y^2} -\left(\frac{1}{y^2}-F(y)\right)dt^2 +\left(\frac{1}{y^2}+F(y)\right)\ell^2d\phi^2
            +2F(y)\, \ell\,  dt\, d\phi \,,
\end{equation}
for the functions $F(y)$ to be specified below. These are exact solutions of PET gravity that can moreover correspond to a Fefferman--Graham expansion of a log black
hole.\footnote{In order to calculate the mass and angular momentum of a black hole, the terms given in an expansion such as
e.g.~\eqref{BTZmetric} are the relevant ones. The solution \eqref{logbhmetric} corresponds exactly to the leading terms
of an expansion of a BTZ or log black hole, depending on the choice of $F(y)$, and gives rise to non-vanishing mass
and angular momentum.}

In case one considers the critical line \eqref{critline}, where either $M_+^2 = 0$ or $M_-^2 = 0$, the function $F(y)$ is given by
\begin{equation}
F(y) = 4G{\rm m} + k\log y \,,
\end{equation}
for some constant $k$. At the tricritical point \eqref{tricritpoint}, where both $M_\pm^2 = 0$, we have
\begin{equation}
F(y) = 4G{\rm m} + k\log y + K \log^2y \,,
\end{equation}
for constants $k$ and $K$. When $k = K = 0$ this solution reduces to \eqref{BTZmetric} with $\rm j=- \ell \rm m$. For $K = 0$, but $k \neq 0$,
we obtain a solution that falls off as $\log y$ towards the $\mathrm{AdS}_3$ boundary. We will refer to this solution as
the `log black hole'. For $K \neq 0$, we obtain a `$\log^2$ black hole', that falls off as $\log^2 y$ towards the
boundary. The masses and angular momenta of these log and $\log^2$ black holes can be calculated using the boundary
stress tensor. The result for the log black hole is
\begin{align}
 \ell M_{\rm log~black~hole}=-J_{\rm log~black~hole}= \frac{3k\,\big(2\sigma+b_2/\ell^4\big)}{G}\,,
\end{align}
while for the $\log^2$ black hole, we obtain
\begin{align}
 \ell M_{\rm log^2~black~hole}=-J_{\rm log^2~black~hole}= \frac{7K\sigma}{G}\,.
\end{align}
Note that on the critical line \eqref{critline}, the mass and angular momentum of the extremal BTZ black hole is zero,
whereas the log black hole has non-zero mass and angular momentum. In that case there is no $\log^2$ black hole solution.
At the tricritical point \eqref{tricritpoint}, both the BTZ and log black holes have zero mass and angular momentum,
whereas the $\log^2$ black hole, present at that point, has non-zero mass and angular momentum. Black holes at critical
lines and points are thus characterized by $\log^n(y)$ asymptotic behavior, where $n$ is a natural number (including $n=0$).
The black holes with the highest possible $n$-value have non-zero mass and angular momentum, whereas the black holes with
lower values of $n$ have zero mass and angular momentum. We expect that this is a general feature of gravity models dual to
higher-rank LCFTs.

\section{Logarithmic modes and dual rank-3 log CFT interpretation} \label{PETCFTsection}

Away from the tricritical point, the six-derivative action we considered in the previous section propagates
one massless and two massive gravitons. At the critical point, the two massive gravitons degenerate with the
massless one and are replaced by new solutions. In contrast to the massless graviton modes, that obey Brown--Henneaux
boundary conditions, these new solutions exhibit $\log$ and $\log^2$ behavior towards the $\mathrm{AdS}_3$ boundary
and are referred to as $\log$ and $\log^2$ modes. The existence of these various logarithmic modes naturally leads
to the conjecture that PET Gravity is dual to a rank-3 logarithmic CFT. In this section, we will discuss these modes
and their AdS/CFT consequences in more detail. We will start by giving explicit expressions for the various modes
at the tricritical point. We will give the boundary stress tensor and use it to evaluate the central
charges of the dual CFT at the tricritical point. Finally, we will comment on the structure of the two-point functions
of the dual CFT at the tricritical point. For the calculation of the on-shell energy of the massless, massive, log and log$^2$ solutions presented in this section we refer the reader to appendix \ref{Energyapp}.

\subsection{Modes at the tricritical point}
The linearized equations of motion \eqref{lineomgauge} can be solved  with the group theoretical approach of \cite{Li:2008dq}. We work in global coordinates, in which the AdS metric is given by
\begin{equation}
ds^2 = \frac{\ell^2}{4}  \left( - du^2- 2 \cosh(2\rho) dudv - dv^2 + 4 d\rho^2 \right)\,,
\end{equation}
where $u$ and $v$ are light-cone coordinates. The solutions of
\eqref{lineomgauge}  form  representations of the $\mathrm{SL}(2,\mathbb{R}) \times \mathrm{SL}(2,\mathbb{R})$
isometry group of $\mathrm{AdS}_3$. These representations can be built up by acting with raising operators of the
isometry algebra on a primary state. A primary state was found in \cite{Li:2008dq} and is given by
\begin{equation}\label{psisol}
\psi_{\mu\nu} = e^{-ihu-i\bar{h}v} (\cosh(\rho))^{-(h+\bar{h})}\sinh^2(\rho) F_{\mu\nu}(\rho)\,,
\end{equation}
with $F_{\mu\nu}(\rho)$ given by
\begin{equation}
F_{\mu\nu}(\rho) =
\left(
\begin{array}{ccc}
 \frac{ h - \bar{h}}{4}+\frac{1}{2} & 0 &  \frac{i \left((h-\bar{h})+2\right)}{4 \cosh \rho \sinh \rho} \\
 0 & \frac{1}{2} -\frac{ h - \bar{h}}{4} &\frac{ i \left(2 -(h - \bar{h})\right)}{4 \cosh \rho \sinh \rho}  \\
 \frac{i \left((h-\bar{h})+2\right)}{4 \cosh \rho \sinh \rho} &
\frac{ i \left(2 -(h - \bar{h})\right)}{4 \cosh \rho \sinh \rho}&
 \frac{ -1}{\cosh^2 \rho \sinh^2 \rho} 
\end{array}
\right) \,.
\end{equation}
The constant weights $h$, $\bar{h}$ obey $h - \bar{h} = \pm 2$, as well as the equation
\begin{align}\begin{split}
& \left( h(h-1) + \bar{h}(\bar{h}-1) -2 \right) \left(2 h(h-1) + 2\bar{h}(\bar{h}-1) -4 -\ell^2 M_+^2 \right)\\ 
& \qquad\qquad \times \left(2 h(h-1) + 2\bar{h}(\bar{h}-1) -4 - \ell^2 M_-^2 \right) =0 \,.
\end{split}\end{align}
This equation has three branches of solutions, corresponding to the massless mode and the two massive modes.
The massless modes obey $\left( h(h-1) + \bar{h}(\bar{h}-1) -2 \right)=0$. The weights which satisfy this equation
and lead to normalizable modes are $(h,\bar{h}) =(2,0)$ and $(0,2)$. They are solutions of linearized Einstein gravity
in AdS$_3$ and correspond to left- and right-moving massless gravitons.

The weights of the other two branches obey $\left(2 h(h-1) + 2\bar{h}(\bar{h}-1) -4 - \ell^2 M_{\pm}^2 \right) =0$.
For those primaries that do not blow up at the boundary $\rho \rightarrow \infty$, we obtain
the following weights:
\begin{align}
\textrm{left-moving}:\; h & = \frac32 +  \frac12\sqrt{1+\ell^2 M_{\pm}^2} \,,
 \qquad\quad \bar{h} = -\frac12 + \frac12 \sqrt{1+\ell^2 M^2_{\pm}} \,, \label{hleft}\\ 
\textrm{right-moving}: \; h & = -\frac12 + \frac12 \sqrt{1+\ell^2 M^2_{\pm}}\,,
\qquad\, \bar{h} = \frac32 +  \frac12\sqrt{1+\ell^2 M_{\pm}^2}\,. \label{hright}
\end{align}
These weights correspond to left- and right-moving massive gravitons, with mass $M_\pm$. The condition that these modes
are normalizable implies that the masses of the modes are real, $M_{\pm}^2 \geq 0$.

At the tricritical point $M_\pm^2 = 0$ and the weights (and therefore the solutions) of the massive modes
degenerate with those of the massless modes. Like in tricritical GMG, there are new solutions, called $\mathrm{log}$
and $\mathrm{log}^2$ modes. Denoting these modes by $\psi^\mathrm{log}$ and $\psi^{\mathrm{log}^2}$ resp., they satisfy
\begin{align}
\cG_{\mu\nu}(\cG(\psi^{\log})) & = 0\, & \textrm{but} & \; \; \; \cG_{\mu\nu}(\psi^{\log}) \neq 0\,, \\
\cG_{\mu\nu}(\cG(\cG(\psi^{\log^2}))) & = 0\, & \textrm{but} & \; \; \; \cG_{\mu\nu}(\cG(\psi^{\log^2})) \neq 0\,.
\end{align}
As was shown in \cite{Grumiller:2008qz}, the log mode can be obtained by differentiation of the massive mode with
respect to $M_\pm ^2 \ell^2$ and by setting $M_\pm^2 = 0$ afterwards:
 \begin{equation}
   \psi^{\log}_{\mu\nu}  =  \frac{\partial \psi_{\mu\nu}(M_{\pm}^2)}{\partial(M_{\pm}^2\ell^2)}  \bigg|_{M_{\pm}^2 = 0} \,.
 \end{equation}
Here $\psi_{\mu\nu}(M_\pm^2)$ is the explicit solution obtained by filling in the weights $(h,\bar{h})$ corresponding to a
massive graviton, in \eqref{psisol}. The $\mathrm{log}^2$ mode can be obtained in a similar way, by differentiating
twice with respect to $M_\pm ^2 \ell^2$. The resulting modes are explicitly given by
 \begin{align}
\psi_{\mu\nu}^{\log} & = f(u,v,\rho)\, \psi_{\mu\nu}^0 \,, \\
\psi_{\mu\nu}^{\log^2} & = \frac12 f(u,v,\rho)^2 \,\psi_{\mu\nu}^0  \,,
\end{align}
where $\psi_{\mu\nu}^0$ corresponds to a massless graviton mode, obtained by using $(h,\bar{h}) = (2,0)$ or $(0,2)$
in \eqref{psisol} and where
 \begin{equation}
   f(u,v,\rho) =  - \frac{i}{2}(u+v)  -\log(\cosh\rho) \,.
 \end{equation}
Note that the massless, log and $\mathrm{log}^2$ modes all behave differently when approaching the boundary
$\rho \rightarrow \infty$. The massless mode obeys Brown--Henneaux boundary conditions.
In contrast, the log mode shows a linear behavior in $\rho$ when taking the $\rho \to \infty$ limit, whereas the
$\log^2$ mode shows $\rho^2$ behavior in this limit. The three kinds of modes therefore all show different
boundary behavior in $\mathrm{AdS}_3$ and the boundary conditions obeyed by log and $\mathrm{log}^2$ modes are
correspondingly referred to as log and $\mathrm{log}^2$ boundary conditions. 

The log and $\log^2$ modes are not eigenstates of the AdS energy operator $H = L_0 + \bar{L}_0$. Instead they
form a rank-3 Jordan cell with respect to this operator (or similarly, with respect to the Virasoro algebra).
The normalization of the log and $\log^2$ modes has been chosen such that when acting on the modes
$h_{\mu\nu}= \{\psi_{\mu\nu}^0 , \psi^{\log}_{\mu\nu} ,\psi^{\log^2}_{\mu\nu} \} $ with $H$, the
off-diagonal elements in the Jordan cell are 1:
\begin{equation} \label{nondecompH}
H \, h_{\mu\nu} = \left( \begin{array}{ccc}
(h+\bar{h}) 	&	0		&	0	\\
	1		& (h+\bar{h})	&	0	\\
	0		&	1		&	(h+\bar{h}) 	
\end{array}
\right) h_{\mu\nu} \,.
\end{equation}
The presence of the Jordan cell shows that the states form indecomposable but non-irreducible representations
of the Virasoro algebra. Furthermore, we have that
\begin{equation} \label{qprim}
L_1 \psi_{\mu\nu}^{\log} = 0 = \bar{L}_1 \psi_{\mu\nu}^{\log} \,, \qquad 
L_1\psi_{\mu\nu}^{\log^2} = 0 = \bar{L}_1 \psi_{\mu\nu}^{\log^2} \,.
\end{equation}
These properties form the basis for the conjecture that PET Gravity is dual to a rank-3 LCFT. The modes correspond
to states in the LCFT and \eqref{nondecompH} translates to the statement that the LCFT Hamiltonian is non-diagonalizable
and that the states form a rank-3 Jordan cell. The conditions \eqref{nondecompH} and \eqref{qprim} indicate that the states associated to
$\psi^{\log}_{\mu\nu}$ and $\psi^{\log^2}_{\mu \nu}$ are quasi-primary. The only proper primary state is the one
associated to $\psi^0_{\mu \nu}$.


\subsection{Boundary stress tensor and structure of the dual CFT}\label{stresstenssec}
To learn more about the dual LCFT, we calculate the boundary stress tensor \cite{Brown:1992br,Kraus:2005zm} of PET gravity. In particular,
we extract from it the central charges. Since the calculation itself is rather technical and not very illuminating
we refer the reader to appendix \ref{stresstensorapp} for details. Here we will give the result for the boundary
stress tensor for PET gravity: 
\begin{align}
 16 \pi G\,T^{\rm 3critgrav}_{ij}=\Big(2\sigma +\frac{\beta}{\ell^2} +\frac{3b_2}{\ell^4}\Big)\gamma_{ij}^{(2)}
                 -\Big(2\sigma +\frac{\beta}{\ell^2}\Big) \gamma_{ij}^{(0)}\gamma_{kl}^{(2)}\gamma^{kl}_{(0)} \,,
\end{align}
where $\gamma_{ij}^{(0)}$,  $\gamma_{ij}^{(2)}$ are the leading, resp.~sub-leading terms in the Fefferman--Graham
expansion of the metric:
\begin{align}\label{poincaremetric}
 ds^2=\frac{dy^2}{y^2}+\gamma_{ij}d x^id x^j, \quad \gamma_{ij}=\frac{1}{y^2}\gamma_{ij}^{(0)}+\gamma_{ij}^{(2)} \,.
\end{align}
Note that we switched to Poincar\'{e} coordinates for convenience.
The central charges follow from the stress tensor \cite{Balasubramanian:1999re} and are given by 
\begin{align}\label{charge3crit}
 c=c_{L/R}=\frac{3\ell}{4G}\,\Big(2\sigma +\frac{\beta}{\ell^2} +\frac{3b_2}{\ell^4} \Big) \,.
\end{align}
 The central charges vanish at the tricritical point, where $\beta= 4\sigma\ell^2$ and $b_2= - 2\sigma\ell^4$.
This lends further support for the conjecture that the dual CFT is logarithmic. Indeed, as  unitary $c=0$ CFTs have no
non-trivial representations, CFTs with central charge $c=0$ are typically non-unitary and thus possibly logarithmic.

The central charges also vanish on the rest of the critical
line \eqref{critline} where just one of the massive modes
becomes massless. On this critical line, the dual CFTs are still expected
to be logarithmic, but the rank must decrease by one with respect to the tricritical point.  The dual theory on the critical
line is thus expected to be an LCFT of rank 2. As a consistency check, we note that (non-critical) NMG is contained in our
model in the limit $b_2\to0$ and $\beta\to1/m^2$. Substituting these values in \eqref{charge3crit}, we see that the central
charge agrees with the central charge found for NMG in \cite{Bergshoeff:2009aq}. 

The dual CFT of PET gravity is thus conjectured to be a rank-3 LCFT with $c_L = c_R = 0$. In that case the general structure
of the two-point correlators is known. The two-point functions are determined by quantities called
new anomalies. If one knows the central charges, one can employ a short-cut \cite{Grumiller:2010tj} to derive these new
anomalies. We do this for the left-moving sector. Similar results hold for the right-moving sector.

Let us start from the non-critical case, where the correlators of the left-moving components $\mathcal{O}^L(z)$ of the boundary
stress tensor are given by
\begin{align}
 \langle \mathcal{O}^L(z) \,\mathcal{O}^L(0) \rangle = \frac{c_L}{2 z^4} \,,
\end{align}
where $c_L$ is given by \eqref{charge3crit}. It may be rewritten in terms of the masses $M_\pm$ as
\begin{align}
 c_L&= \frac{3\ell^3\sigma}{G}\frac{M_-^2M_+^2}{M_-^2+M_+^2+\frac{1}{\ell^2}+2\ell^2M_-^2M_+^2} \,.
\end{align}
Let us first consider the case where only one of the two masses vanishes, e.g.~when $M_+^2 \rightarrow 0$. In this case,
we are on the critical line \eqref{critline}. The CFT dual is conjectured to be a rank-2 LCFT with vanishing central charges.
The two-point functions for such an LCFT are of the form
\begin{subequations}\begin{align}
 \langle \mathcal{O}^L(z) \,\mathcal{O}^L(0) \rangle &= 0 \,, \\
 \langle \mathcal{O}^L(z) \,\mathcal{O}^{\rm log}(0) \rangle &= \frac{b_L}{2 z^4} \,, \\
 \langle \mathcal{O}^{\rm log}(z,\bar{z}) \,\mathcal{O}^{\rm log}(0) \rangle &= -\frac{b_L\,\log|z|^2}{ z^4} \,,
\end{align}\end{subequations}
where $\mathcal{O}^\mathrm{log}(z,\bar{z})$ denotes the logarithmic partner of $\mathcal{O}^L(z)$.
The parameter $b_L$ is the new anomaly. 
It can be calculated from the central charges and the difference of the conformal weights of the left-moving primary, $(h,\bar{h})=(2,0)$, and the left-moving massive
mode, see equation \eqref{hleft}, via a limit procedure.
This difference, up to linear order for small $M_+^2$, is given by
\begin{align}
 2\Delta_{LM_+} &\equiv2(h_L-h_{M_+})=\big(1-\sqrt{1+\ell^2M_+^2}\big) \approx-\frac{\ell^2M_+^2}{2} \,.
\end{align}
The new anomaly is then given by
\begin{align}
	b_L&= \lim_{\ell^2M_+^2\to0} \frac{c_L}{\Delta_{LM_+}} \\
 &= \lim_{\ell^2M_+^2\to0}\,-\frac{4}{\ell^2M_+^2}\frac{3\ell^3\sigma}{G}
                           \frac{M_-^2M_+^2}{M_-^2+M_+^2+\frac{1}{\ell^2}+2\ell^2M_-^2M_+^2}
    =-\frac{12\ell\sigma}{G}\frac{M_-^2}{M_-^2+\frac{1}{\ell^2}} \,. \label{bL}
\end{align}
Note that in the limit $b_2\to0$, where we recover critical NMG, we find $M_-^2\to\infty$ and the corresponding limit
of the result \eqref{bL} agrees with the new anomaly of NMG \cite{Grumiller:2009sn}.

At the tricritical point, the correlators are conjectured to be the ones of a rank-3 LCFT with vanishing central charges:
\begin{subequations}\label{2ptlcft}\begin{align}
 \langle \mathcal{O}^L(z) \,\mathcal{O}^L(0) \rangle = \langle \mathcal{O}^L(z) \,\mathcal{O}^{\rm log}(0) \rangle &=0 \,, \\
 \langle \mathcal{O}^{L}(z) \,\mathcal{O}^{\rm log^2}(0) \rangle = \langle \mathcal{O}^{\rm log}(z) \,\mathcal{O}^{\rm log}(0) \rangle &= \frac{a_L}{2 z^4}  \,, \\
 \langle \mathcal{O}^{\rm log}(z,\bar{z}) \,\mathcal{O}^{\rm log^2}(0) \rangle &= -\frac{a_L\,\log|z|^2}{z^4} \,, \\
 \langle \mathcal{O}^{\rm log^2}(z,\bar{z}) \,\mathcal{O}^{\rm log^2}(0) \rangle &= \frac{a_L\,\log^2|z|^2}{z^4}\,.
\end{align}\end{subequations}
Here $\mathcal{O}^{\rm log}(z,\bar{z})$, $\mathcal{O}^{\rm log^2}(z,\bar{z})$ are the two logarithmic partners of $\mathcal{O}^L(z)$.
The new anomaly $a_L$ at the tricritical point is obtained via another limit:
\begin{align}
	a_L=\lim_{\ell^2M_-^2\to0} \frac{b_L}{\Delta_{LM_-}} = \lim_{\ell^2M_-^2\to0}\,\frac{4}{\ell^2M_-^2}\frac{12\ell\sigma}{G}\frac{M_-^2}{M_-^2+\frac{1}{\ell^2}}
    =\frac{48\ell\sigma}{G}\,.
\end{align}
Knowledge of the central charges thus allows one to obtain the new anomalies and hence fix the structure of the
two-point correlators, via the limit procedure of \cite{Grumiller:2010tj}.

\section{Truncation of PET gravity}  \label{truncsection}

In the previous sections, we discussed the six-derivative PET gravity model and showed that the linearized theory has
solutions that obey Brown--Henneaux, log and $\mathrm{log}^2$ boundary conditions. This led to the conjecture that three-dimensional PET gravity,
with boundary conditions that include all these solutions, is dual to a rank-3 LCFT.  We have
calculated the central charges and new anomalies of these conjectured LCFTs.

In this section we will consider a truncation of PET gravity, that is defined by only keeping modes that obey
Brown--Henneaux and log boundary conditions. The same truncation was considered in the context of a scalar field toy model in \cite{Bergshoeff:2012sc}. In this section, we will show that this truncation, phrased
as a restriction of the boundary conditions, is equivalent to considering a sub-sector of the theory, that has
zero values for the conserved Abbott--Deser--Tekin charges associated to (asymptotic) symmetries \cite{Deser:2002jk,Deser:2002rt}.
We will start by calculating these conserved charges. After that, we will evaluate them for generic solutions of
the non-linear theory, that obey Brown--Henneaux, log or $\log^2$ boundary conditions. We will comment on the
truncation afterwards.


In order to calculate the conserved Killing charges we will follow the line of reasoning proposed in
\cite{Deser:2002jk,Deser:2002rt} (later worked out for the log modes in NMG in \cite{Liu:2009kc}). When the
background admits a Killing vector $\xi_{\mu}$, one can define a covariantly conserved current by
\begin{equation}
 \cK^{\mu} =  \xi_{\nu} T^{\mu\nu}\,,
\end{equation}
with $T_{\mu\nu}$ the conserved energy-momentum tensor. This energy-momentum tensor is defined by considering a split
of the metric $g_{\mu \nu}$ in a background metric $\bar{g}_{\mu \nu}$ (that solves the vacuum field equations) and a
(not necessarily infinitesimal) deviation $h_{\mu \nu}$
\begin{equation}
g_{\mu \nu} = \bar{g}_{\mu \nu} + h_{\mu \nu} \,.
\end{equation}
The field equations can then be separated in a part that is linear in $h_{\mu \nu}$ and a part that contains all
interactions. The latter, together with a possible stress tensor for matter, constitute $T_{\mu \nu}$. 
The full field equations, written in terms of $h_{\mu \nu}$ take the form
\begin{equation}
\mathcal{E}_{\mu \nu}{}^{\alpha \beta} h_{\alpha \beta} = T_{\mu \nu} \,, \label{epshT}
\end{equation}
where $\mathcal{E}_{\mu \nu}{}^{\alpha \beta}$ is a linear differential operator acting on $h_{\mu \nu}$. 
The energy-momentum tensor $T_{\mu \nu}$ is given by the left hand side of equation \eqref{epshT}, i.e.~the part of the
field equations linear in $h_{\mu \nu}$ and this was found in section \ref{PETsection}.

 Since the current $\cK^{\mu}$ is covariantly conserved,
there exists an anti-symmetric two-form $\cF^{\mu\nu}$, such that $ \xi_{\nu} T^{\mu\nu} = \bnabla_{\nu} \cF^{\mu\nu}$.
The conserved Killing charges can then be expressed as
\begin{equation}\label{charges}
Q^{\mu} =  \int_{\cM} d^2 x \sqrt{-\bg}T^{\mu\nu}\xi_{\nu} =  \int_{\partial\cM} dS_i \sqrt{-\bg} \cF^{\mu i}\,,
\end{equation}
where $\cM$ is a spatial surface with boundary $\partial\cM$. We may find the expression for $\cF^{\mu\nu}$ for PET gravity by
writing the linearized equations of motion \eqref{lineom} contracted with a Killing vector as a total derivative. The
first term in the first line of \eqref{lineom} is  the linearized Einstein tensor, which may be written as
\begin{equation}\label{KillingG}
\xi_{\mu} \cG^{\mu\nu}(h) = \bnabla_{\rho} \left\{ \xi_{\nu} \bnabla^{[\mu}h^{\rho] \nu} + \xi^{[\mu} \bnabla^{\rho ]} h + h^{\nu [ \mu} \bnabla^{\rho]}\xi_{\nu} - \xi^{[\mu}\bnabla_{\nu} h^{\rho] \nu} + \frac12 h \bnabla^{\mu}\xi^{\rho} \right\}\,.
\end{equation}
In the second term in the first line of \eqref{lineom}, we may replace $h_{\mu\nu}$ in the above expression with $\cG_{\mu\nu}(h)$
to obtain
\begin{align}\label{KillingGG}
\xi_{\mu} \cG^{\mu\nu}(\cG(h)) = & \; \bnabla_{\rho} \bigg\{ \xi_{\nu} \bnabla^{[\mu}\cG^{\rho] \nu}(h) + \xi^{[\mu} \bnabla^{\rho ]} \cG(h) + \cG^{\nu [ \mu}(h) \bnabla^{\rho]}\xi_{\nu} + \frac12 \cG(h) \bnabla^{\mu}\xi^{\rho} \bigg\}\,,
\end{align}
where we denoted $ \bg^{\mu\nu} \cG_{\mu\nu}(h) = \cG(h)$. The same trick can be used to calculate the
$\xi_{\nu}\cG^{\mu\nu}(\cG(\cG(h)))$-term. It is given by
\begin{align}\label{GGGh}
\xi_{\mu} \cG^{\mu\nu}(\cG(\cG(h))) = & \; \bnabla_{\rho} \bigg\{ \xi_{\nu} \bnabla^{[\mu}\cG^{\rho] \nu}(\cG(h)) + \xi^{[\mu} \bnabla^{\rho ]} \cG(\cG(h)) + \cG^{\nu [ \mu}(\cG(h)) \bnabla^{\rho]}\xi_{\nu} \\ 
 \nonumber &  + \frac12 \cG(\cG(h)) \bnabla^{\mu}\xi^{\rho} \bigg\}\,. 
\end{align}
The last two lines in \eqref{lineom} are given by
\begin{align}
\xi_{\nu} \left(\bg^{\mu\nu} \bBox - \bnabla^{\mu} \bnabla^{\nu} - \frac{2}{\ell^2}\bg^{\mu\nu}\right) R^{(1)} & =
 - 4 \bnabla_{\rho}\left(  \xi^{[\mu}\bnabla^{\rho]} \cG(h) +\frac12 \cG(h) \bnabla^{\mu} \xi^{\rho} \right) \,,\\
\xi_{\nu} \left(\bg^{\mu\nu} \bBox - \bnabla^{\mu} \bnabla^{\nu} - \frac{2}{\ell^2}\bg^{\mu\nu}\right) \bBox R^{(1)} & =
 - 4 \bnabla_{\rho}\left(  \xi^{[\mu}\bnabla^{\rho]} \bBox\cG(h) +\frac12 \bBox\cG(h) \bnabla^{\mu} \xi^{\rho} \right) \,.
\end{align}
Combining all this, we may write
\begin{align}\begin{split}\label{Tmunu}
16 \pi G \xi_{\mu} T^{\mu\nu} =& \; \bar{\sigma} \xi_{\nu}\cG^{\mu\nu}(h) - \frac{1}{\ell^2}(2\beta \ell^2 + 4 b_2) \xi_{\nu} \cG^{\mu\nu}(\cG(h)) -4b_2\xi_{\mu} \cG^{\mu\nu}(\cG(\cG(h)))   \\
& + \frac{1}{\ell^2}( b_2 + \beta \ell^2) \bnabla_{\rho}\left(  \xi^{[\mu}\bnabla^{\rho]} \cG(h) +\frac12 \cG(h) \bnabla^{\mu} \xi^{\rho} \right) \\
& + b_2 \bnabla_{\rho}\left(  \xi^{[\mu}\bnabla^{\rho]} \bBox\cG(h) +\frac12 \bBox\cG(h) \bnabla^{\mu} \xi^{\rho} \right) \\
= & \; 16 \pi G \bnabla_{\rho}\cF^{\mu\rho} \,.
\end{split}\end{align}
The first three terms in this expression are given by \eqref{KillingG}, \eqref{KillingGG} and \eqref{GGGh}. We are now ready to
specify the boundary conditions and explicitly calculate the charges.

\subsection{Brown--Henneaux boundary conditions}

Let us first show that demanding the deviations $h_{\mu\nu}$ respect Brown--Henneaux boundary conditions leads to finite charges.
In order to simplify the calculation,
we will work with the AdS$_3$ metric in the Poincar\'e patch. Using light cone coordinates, this metric is given by
\begin{equation}
ds^2 = \frac{\ell^2}{r^2} dr^2 - \frac{\ell^2}{r^2}dx^+dx^- \,,
\end{equation}
where the conformal boundary is at $r\to0$.
We now consider deviations $h_{\mu \nu}$ that fall off according to Brown--Henneaux boundary conditions, i.e.~the metric must behave as:
\begin{align}\label{BHbc}
g_{+-} & = -\frac{\ell^2}{2r^2} + \cO(1)\,, & g_{++} & = \cO(1)\,, & g_{--} & = \cO(1) \,, \\
g_{rr} & = \frac{\ell^2}{r^2} + \cO(1)\,, & g_{+r} & = \cO(r)\,, & g_{-r} & = \cO(r)\,. \nonumber
\end{align}
The most general diffeomorphisms that preserve the asymptotic form of the metric are generated by asymptotic Killing vectors
$\xi$ that are explicitly given by
\begin{align}\begin{split}
\xi  = & \; \xi^{+}\partial_{+} + \xi^{-}\partial_{-} + \xi^{r} \partial_{r}  \\ 
 = &\;  \left(\epsilon^{+}(x^+) + \frac{r^2}{2} \partial_-^2\epsilon^-(x^-) + \cO(r^4) \right)\partial_+ \\
 & + \left(\epsilon^{-}(x^-) + \frac{r^2}{2} \partial_+^2\epsilon^+(x^+) + \cO(r^4) \right)\partial_- \\
& + \frac12 r \left(\partial_+ \epsilon^{+}(x^+) + \partial_- \epsilon^{-}(x^-) + \cO(r^3)  \right) \partial_{r} \,. 
\end{split}\end{align}
By choosing the basis  such that
\begin{equation}
\xi^L_{m} = \xi(\epsilon^+ = 0, \epsilon^-= e^{imx^-}) \,, \qquad  \xi^R_{m} = \xi(\epsilon^+ =  e^{imx^+}, \epsilon^-= 0)\,, 
\end{equation}
one can see that the asymptotic symmetry algebra, generated by the asymptotic Killing vectors, is given by two copies of
the Virasoro algebra on the boundary. 

We now parametrize the deviations $h_{\mu\nu}$ in terms of functions $f_{\mu\nu}(x^+,x^-)$ such that the Brown--Henneaux
boundary conditions \eqref{BHbc} are satisfied:
\begin{align}\label{paraBHbc}
h_{--} & =  f_{--}(x^+,x^-) + \ldots \,, & \nonumber
h_{++} & =  f_{++}(x^+,x^-) + \ldots \,, \\
h_{+-} & = f_{+-}(x^+,x^-) + \ldots \,, &
h_{rr} & = f_{rr}(x^+,x^-) + \ldots \,, \\ 
h_{+r} & = r f_{+r}(x^+,x^-) + \ldots \,, & 
h_{-r} & = r f_{-r}(x^+,x^-) + \ldots \,. \nonumber
\end{align}
Here the dots denote sub-leading terms which vanish more quickly as we move towards the boundary of AdS$_3$. They
do not contribute to the conserved charges. 

The conserved charge is calculated by
\begin{equation}\label{CCpoincare}
Q =  \lim_{r \to 0} \frac{1}{16 \pi G} \int \extd\phi \,\sqrt{-\bg}\, \cF^{t r}\,,
\end{equation} 
where $\phi = \tfrac12 (x^+ - x^-)$ and $t = \tfrac12 ( x^+ + x^-)$. Inserting the boundary conditions
\eqref{paraBHbc} into \eqref{Tmunu} we find
\begin{align}\begin{split}\label{QBH}
Q = \frac{1}{16 \pi G \ell} \int \extd\phi \;  & \bigg\{ \left(\sigma + \frac12 \frac{\beta}{\ell^2}
          + \frac32 \frac{b_2}{\ell^4} \right) \Big( \epsilon^+  f_{++} + \epsilon^- f_{--}\Big) \\ 
& - \left(\sigma - \frac12 \frac{\beta}{\ell^2} - \frac32 \frac{b_2}{\ell^4} \right)
         \left( \frac{(\epsilon^+ + \epsilon^- )( 4f_{+-} - f_{rr} )}{4}\right)\bigg\} \,.
\end{split}\end{align}
The $rr$-component of the linearized equations of motion gives the asymptotic constraint
\begin{equation}\label{asymptconstr}
4f_{+-} - f_{rr} = 0\,.
\end{equation}
Using this relation, the left- and right-moving charges of solutions that obey Brown--Henneaux boundary conditions are given by
\begin{align}\label{QLBH}
Q_L & = \frac{1}{16 \pi G \ell} \left(\sigma + \frac12 \frac{\beta}{\ell^2} + \frac32 \frac{b_2}{\ell^4} \right)
                 \int \extd\phi \;   \epsilon^- f_{--} \,,\\ 
Q_R & = \frac{1}{16 \pi G \ell} \left(\sigma + \frac12 \frac{\beta}{\ell^2} + \frac32 \frac{b_2}{\ell^4} \right)
                 \int \extd\phi \;   \epsilon^+ f_{++}  \,. \label{QRBH} 
\end{align}
These charges are always finite for arbitrary values of the parameters.  Note that, at the critical line \eqref{critline}
and at the tricritical point \eqref{tricritpoint}, the coefficients in front of the charges vanish and the charges
are zero. This is analogous to what happened to the mass of the BTZ black hole, calculated via the boundary stress tensor.

\subsection{Log boundary conditions}

Imposing Brown--Henneaux boundary conditions at the critical point only allows for zero charges and does not include solutions with log and $\log^2$
behavior (such as for instance the log and $\log^2$ black holes given in section \ref{BHsection}). In order to
remedy this we need to relax the Brown--Henneaux boundary conditions to include logarithmic behavior in $r$.
We may do so in analogy to the log  boundary conditions in TMG \cite{Grumiller:2008es,Maloney:2009ck}. In
Poincar\'e coordinates, we now require that the metric falls off as:
\begin{align}\label{logbc}
g_{+-} & = -\frac{\ell^2}{2r^2} + \cO(1)\,, & g_{++} & = \cO(\log r)\,, & g_{--} & = \cO(\log r) \,,\\
g_{rr} & = \frac{\ell^2}{r^2} + \cO(1)\,, & g_{+r} & = \cO(r\log r)\,, & g_{-r} & = \cO(r \log r)\,. \nonumber
\end{align}
The asymptotic Killing vector $\xi$ compatible with this new logarithmic behavior only changes in the sub-subleading terms:
\begin{align}\begin{split}
\xi  = & \; \xi^{+}\partial_{+} + \xi^{-}\partial_{-} + \xi^{r} \partial_{r} \\ 
 = &\;  \left(\epsilon^{+}(x^+) + \frac{r^2}{2} \partial_-^2\epsilon^-(x^-) + \cO(r^4\log r) \right)\partial_+ \\
 & + \left(\epsilon^{-}(x^-) + \frac{r^2}{2} \partial_+^2\epsilon^+(x^+) + \cO(r^4 \log r) \right)\partial_- \\
& + \frac12 r \left(\partial_+ \epsilon^{+}(x^+) + \partial_- \epsilon^{-}(x^-) + \cO(r^3)  \right) \partial_{r} \,.
\end{split}\end{align}
This ensures that the Virasoro algebra of the asymptotic symmetry group is preserved.
We parametrize the metric deviations consistent with the boundary conditions \eqref{logbc} as:
\begin{align}\label{paralogbc}
h_{--} & = \log r f_{--}^{\log}(x^+,x^-) + \ldots \,, & \nonumber
h_{++} & = \log r f_{++}^{\log}(x^+,x^-) + \ldots \,, \\
h_{+-} & = f_{+-}^{\log}(x^+,x^-) + \ldots \,, &
h_{rr} & = f_{rr}^{\log}(x^+,x^-) + \ldots \,, \\ 
h_{+r} & = r\log r f_{+r}^{\log}(x^+,x^-) + \ldots \,, & 
h_{-r} & = r \log r f_{-r}^{\log}(x^+,x^-) + \ldots \,. \nonumber
\end{align}
Imposing the asymptotic constraint from the $rr$-component of the equations of motion \eqref{asymptconstr}, we find
that the charge, subject to log boundary conditions, is
\begin{align}\begin{split}\label{Qlogbc}
Q^{\mathrm{log}} = \frac{1}{16 \pi G\ell} \int\extd\phi \;  & \bigg\{ \left(\sigma + \frac12 \frac{\beta}{\ell^2}
     + \frac32 \frac{b_2}{\ell^4} \right) \log(r) \left( \epsilon^+  f_{++}^{\log} + \epsilon^- f_{--}^{\log}\right) \\ 
& + \left(\frac12 \sigma + \frac{9}{4} \frac{\beta}{\ell^2} + \frac{19}{4} \frac{b_2}{\ell^4} \right)
    \left( \epsilon^+  f_{++}^{\log} + \epsilon^- f_{--}^{\log}\right)\bigg\} \,.
\end{split}\end{align}
This expression diverges except at the critical line \eqref{critline}. Note that only on this line solutions with
asymptotic log behavior are expected and thus log boundary conditions only make sense here. At the tricritical
point, we find that both left and right charges vanish:
\begin{equation}
Q_L^{\log}  = 0\,, \qquad {\rm and} \qquad Q_R^{\log}  = 0\,.
\end{equation}
This is again analogous to what happened to the mass of the log black hole at the tricritical point.

\subsection{Log$^2$ boundary conditions}

We may go one step beyond log  boundary conditions and allow for $\log^2$ behavior. This is similar to the boundary conditions
imposed in \cite{Liu:2009pha} for the left-moving modes at the tricritical point in GMG, the only difference being that now we
deal with a parity even theory, so both left- and right-moving sectors need to be relaxed.  
Transferring the $\log^2$ boundary conditions of \cite{Liu:2009pha} to the Poincar\'e patch of AdS$_3$ and including
similar conditions for the right-movers we find that the asymptotic behavior of the metric should be:
\begin{align}\label{log2bc}
g_{+-} & = -\frac{\ell^2}{2r^2} + \cO(1)\,, & g_{++} & = \cO(\log^2 r)\,, & g_{--} & = \cO(\log^2 r) \,,\\
g_{rr} & = \frac{\ell^2}{r^2} + \cO(1)\,, & g_{+r} & = \cO(r\log^2 r)\,, & g_{-r} & = \cO(r \log^2 r)\,. \nonumber
\end{align}
Again the asymptotic Killing vector $\xi$ only changes in the sub-subleading term:
\begin{align}\begin{split}
\xi  = & \; \xi^{+}\partial_{+} + \xi^{-}\partial_{-} + \xi^{r} \partial_{r}  \\ 
 = &\;  \left(\epsilon^{+}(x^+) + \frac{r^2}{2} \partial_-^2\epsilon^-(x^-) + \cO(r^4\log^2 r) \right)\partial_+ \\
 & + \left(\epsilon^{-}(x^-) + \frac{r^2}{2} \partial_+^2\epsilon^+(x^+) + \cO(r^4 \log^2 r) \right)\partial_-  \\
& + \frac12 r \left(\partial_+ \epsilon^{+}(x^+) + \partial_- \epsilon^{-}(x^-) + \cO(r^3)  \right) \partial_{r} \,,
\end{split}\end{align}
and the Virasoro algebra of the asymptotic symmetry group is preserved. The
deviations $h_{\mu\nu}$ are parametrized as:
\begin{align}\label{paralog2bc}
h_{--} & = \log^2 r f_{--}^{\log^2}(x^+,x^-) + \ldots \,, & \nonumber
h_{++} & = \log^2 r f_{++}^{\log^2}(x^+,x^-) + \ldots \,, \\
h_{+-} & = f_{+-}^{\log^2}(x^+,x^-) + \ldots \,, &
h_{rr} & = f_{rr}^{\log^2}(x^+,x^-) + \ldots \,, \\ 
h_{+r} & = r\log^2 r f_{+r}^{\log^2}(x^+,x^-) + \ldots \,, & 
h_{-r} & = r \log^2 r f_{-r}^{\log^2}(x^+,x^-) + \ldots \,. \nonumber
\end{align}
Computing the conserved charges at the tricritical point \eqref{tricritpoint}
we find
\begin{align}\label{QLlog2}
Q^{\log^2}_L =  \frac{ \sigma}{ \pi G \ell} \int\extd\phi \;  \epsilon^- f_{--}^{\log^2} \,, \\
Q^{\log^2}_R =  \frac{ \sigma}{ \pi G \ell} \int\extd\phi \;  \epsilon^+  f_{++}^{\log^2}   \,, \label{QRlog2}
\end{align}
which is finite. Thus we obtain exactly the same structure that we obtained earlier for the masses
of the BTZ, log and log$^2$ black hole.

\subsection{Truncating PET gravity}

Above, we have indicated that one can define boundary conditions that lead to finite charges at all points in parameter space, that include solutions with possible log and $\mathrm{log}^2$ behavior, if these are present at the point under consideration. At the tricritical point, the conserved charges for Brown--Henneaux and log boundary conditions vanish while
the charge for modes with $\log^2$ boundary are generically non-zero.

In \cite{Bergshoeff:2012sc} it was suggested that any modes with $\log^2$ behavior towards the boundary can be discarded. Only modes which
satisfy Brown--Henneaux and log boundary conditions are then kept. Since the conserved charges associated with Brown-Henneaux and log
boundary behavior vanish, the truncation of $\log^2$ modes may be rephrased as a restriction to a zero charge sub-sector of the
theory, in analogy to \cite{Maloney:2009ck}.
More precisely, we require that both charges $Q_L$ and $Q_R$, corresponding to the left- and right-moving excitations, vanish independently.
Then --- given that $\epsilon^+$ and $\epsilon^-$ form a complete basis\footnote{This is most easily seen by choosing
$\epsilon^+\sim e^{inx^+}$ and $\epsilon^-\sim e^{inx^-}$.} --- setting the charges given by \eqref{QLlog2} and \eqref{QRlog2} to zero
implies that $f_{++}^{\rm log^2}$ and $f_{--}^{\rm log^2}$ must be zero. Furthermore, there is enough gauge freedom to gauge $f_{+r}^{\rm log^2}$ and $f_{-r}^{\rm log^2}$ away. Hence, requiring $Q_L$ and $Q_R$ to vanish
is equivalent to imposing log boundary conditions. Preservation of the boundary conditions under time evolution can then be rephrased as charge conservation, at the classical level. 

The above argument rephrases the truncation of \cite{Bergshoeff:2012sc} as a restriction to a zero charge sub-sector, at the level of non-linear PET gravity. One can also consider this truncation for the linearized theory. According to \cite{Bergshoeff:2012sc}, applying the truncation to the correlation functions \eqref{2ptlcft} of PET gravity
consists of removing all log$^2$ operators. Subsequently, we find that the remaining two-point
functions contain a non-vanishing result for the log-log correlator:
 \begin{align} \begin{split}\label{trunccorr}
 \langle \mathcal{O}^L(z) \,\mathcal{O}^L(0) \rangle = \langle \mathcal{O}^L(z) \,\mathcal{O}^{\rm log}(0) \rangle &=0 \,, \\
 \langle \mathcal{O}^{\rm log}(z,\bar{z}) \,\mathcal{O}^{\rm log}(0) \rangle &= \frac{a_L}{2 z^4}  \,.
\end{split}\end{align}
The structure of the remaining non-zero correlator is identical to that of an ordinary CFT. This implies that the log
mode in the truncated gravity theory is not a null mode, at least in the linearized theory. Truncating linearized PET gravity by restricting to log boundary conditions may be contrasted
to the truncation of (the higher-dimensional analogue of) critical NMG \cite{Liu:2009bk,Liu:2009kc,Lu:2011zk} by restricting to Brown--Henneaux boundary conditions. Applying the
truncation to these theories does not lead to non-trivial two-point correlators.

Similar conclusions can be reached by calculating the scalar product on the state space of the CFT dual to linearized PET
gravity. This calculation can be done on the gravity side, along the lines of \cite{Porrati:2011ku}. Using the linearized
action at the tricritical point \eqref{L2tricrit}, as well as the ensuing equations of motion \eqref{w4}, \eqref{w5},
\eqref{w6}, we find that the inner product on the state space is given by:
 \begin{align}\begin{split}
	 \bracket{\psi}{\phi} \sim \int \rmd^3 x \sqrt{|\bar g|} \bar g^{00}  & \left\{ 
		(\bBox - 2\Lambda)^2 \psi^{*\mu\nu} {D_0}  \phi_{\mu\nu}  
		+ \psi^{*\mu\nu} {D_0}  (\bBox - 2\Lambda)^2 \phi_{\mu\nu}  		\right.\\		  
		&\left.	  + (\bBox - 2\Lambda)\psi^{*\mu\nu} {D_0}  (\bBox - 2\Lambda)\phi_{\mu\nu} \right\} \,,
\end{split} \end{align}  
where $\ket{\psi}$ denotes the CFT state corresponding to the mode $\psi_{\mu \nu}$. Introducing an index $i$ that can denote
either a massless mode, a log mode or a $\mathrm{log}^2$ mode,
\begin{equation}
\psi^i= \{ \psi^{(0)}, \psi^{\mathrm{log}},  \psi^{\mathrm{log}^2}\} \,,
\end{equation}
the scalar product has the following structure:
 \begin{align} \label{innprodlcft}
	 \bracket{\psi_i}{\phi_j} \sim A_{ij}\,, \qquad A =   
  \begin{pmatrix}
	 0 & 0 & a \\
	 0 & b & c \\
	 a & c & d
 \end{pmatrix} \,,
 \end{align}
 where $a$, $b$, $c$, $d$ are generically non-zero entries. Note that the structure of the inner product is similar to
that of the two-point functions \eqref{2ptlcft}. The inner product \eqref{innprodlcft} is indefinite and negative norm
states can be constructed as linear combinations of modes that have a mutual non-zero inner product. Upon truncating
states that correspond to $\mathrm{log}^2$ modes, the inner product assumes the form:
 \begin{align}
	 \bracket{\psi_i}{\phi_j} \sim A_{ij}\,, \qquad A =   
  \begin{pmatrix}
	 0 & 0 \\
	 0 & b \\ 
 \end{pmatrix}\,, 
 \end{align} 
 where the index $i$ now only corresponds to massless and log modes.
After truncation, the state corresponding to the massless graviton has zero norm, while the log state can have a
non-negative norm. The massless graviton state moreover has no overlap with the log state, so in principle, at the linearized level in which this analysis is done, the truncation could consist of a unitary
sector plus an extra null state, in analogy to what was found in \cite{Bergshoeff:2012sc}.

\section{Conclusion and discussion} \label{conclsection}

In this paper, we have considered three-dimensional, tricritical higher-derivative gravity theories around $\mathrm{AdS}_3$.
These tricritical theories are obtained by considering higher-derivative gravities, that ordinarily propagate one massless
and two massive graviton states, at a special point in their parameter space where all massive gravitons become
massless. The massive graviton solutions, that ordinarily obey Brown--Henneaux boundary conditions, are in tricritical
theories replaced by new solutions that obey $\log$ and $\log^2$ boundary conditions towards the AdS boundary: the so-called
$\log$ and $\log^2$ modes.

GMG at the tricritical point constitutes a parity odd example of such a tricritical gravity theory and was studied in
\cite{Liu:2009pha}. It was also shown that this theory is dual to a parity violating, rank-3 LCFT. In this paper, we
constructed a parity even example, called Parity Even Tricritical (PET) gravity, that is of sixth order in derivatives.
We have given explicit expressions for the $\log$ and $\log^2$ modes and we have given indications that the existence
of these modes leads to the conjecture that PET gravity is dual to a rank-3 LCFT. We have calculated the central charges
and new anomaly of this LCFT via a calculation of the boundary stress tensor. We have also calculated the conserved
charges, associated to the (asymptotic) symmetries, and found that, at the critical point, these charges vanish for
excitations that obey Brown--Henneaux and $\log$ boundary conditions, whereas they are generically non-zero for states
with $\log^2$ boundary conditions.

The results of this paper can be put in the context of the findings of \cite{Bergshoeff:2012sc}, where a scalar field model was studied,
that (in the six-derivative case) can be seen as a toy model for PET Gravity. There it was argued that odd rank LCFTs
allow for a non-trivial truncation, that on the gravity side can be seen as restricting oneself to Brown--Henneaux and
$\log$ boundary conditions. Here, we found that similar conclusions hold for PET gravity, at the linearized level. Indeed, upon applying this truncation to the two-point correlators of the dual LCFT,  the truncated theory still has one non-trivial correlator.

In order to go beyond the linearized level, one should first address the issue of the consistency of the truncation, in the presence of interactions. In this paper we have made a step in this direction by rephrasing the truncation for PET gravity as restricting oneself to a zero
charge sub-sector of the theory, with respect to the Abbott--Deser--Tekin charges associated to (asymptotic) symmetries. Similar
conclusions can be made for tricritical GMG using the results for the conserved charges in \cite{Liu:2009pha}. This
reformulation of the truncation of \cite{Bergshoeff:2012sc} can
be useful for showing the consistency of the truncation at the non-linear level. Indeed, classically, the
consistency of the truncation follows from charge conservation \cite{Maloney:2009ck}.
It is conceivable, however, that log$^2$-modes are generated in higher-order correlation
functions \cite{Skenderis:2009kd}. Thus, the calculation of these correlators is needed in order to be able to say more about the consistency of the truncation, beyond the classical level.

In case the consistency of the truncation could be rigorously proven, an interesting question regards the meaning of the
truncated theory. The truncated theory would be conjectured to have two-point functions given by eqs.~\eqref{trunccorr}.
It is unclear what CFT could give rise to such two-point functions. Moreover, naively the truncation does not affect
the value of the central charge. The truncated theory would still seem to have zero central charges and hence
be either non-unitary or trivial. Moreover, it is unclear whether zero charge bulk excitations can account for the apparent non-triviality of the truncated LCFT . It thus seems very hard to obtain non-trivial, unitary truncations and it is likely that the apparent non-triviality is an artifact of the linearized approximation. It is  however
useful to note that the correlator of two logarithmic operators $\mathcal{O}^{\rm log}$ has the form of a two-point
function of the left-moving components of the energy-momentum tensor of an ordinary CFT with central charge given by
the new anomaly $a_L$. It thus seems that the logarithmic operators play the role of the energy-momentum tensor in the
truncated theory. It remains to be seen whether this is the case and can possibly lead to a unitary theory.

\subsection*{Acknowledgments}
The authors wish to thank Niklas Johansson, Teake Nutma and Massimo Porrati for useful discussions and the organizers of the Carg\`{e}se Summer School for providing the pleasant and stimulating environment, where this work was completed. SdH, WM and JR are supported by the
Dutch stichting voor Fundamenteel Onderzoek der Materie (FOM). The authors acknowledge use of the xAct  software package.

\appendix

\section{Truncating tricritical GMG} \label{GMGapp}

In this section, we will consider the truncation procedure of \cite{Bergshoeff:2012sc} in the context of tricritical GMG.
The main results that are necessary for this discussion were obtained in \cite{Liu:2009pha}. In this appendix, we will
collect these results and interpret them in terms of the truncation procedure.

\subsection{GMG and its tricritical points} 

\begin{figure}
\centering
\includegraphics[width=0.8\textwidth]{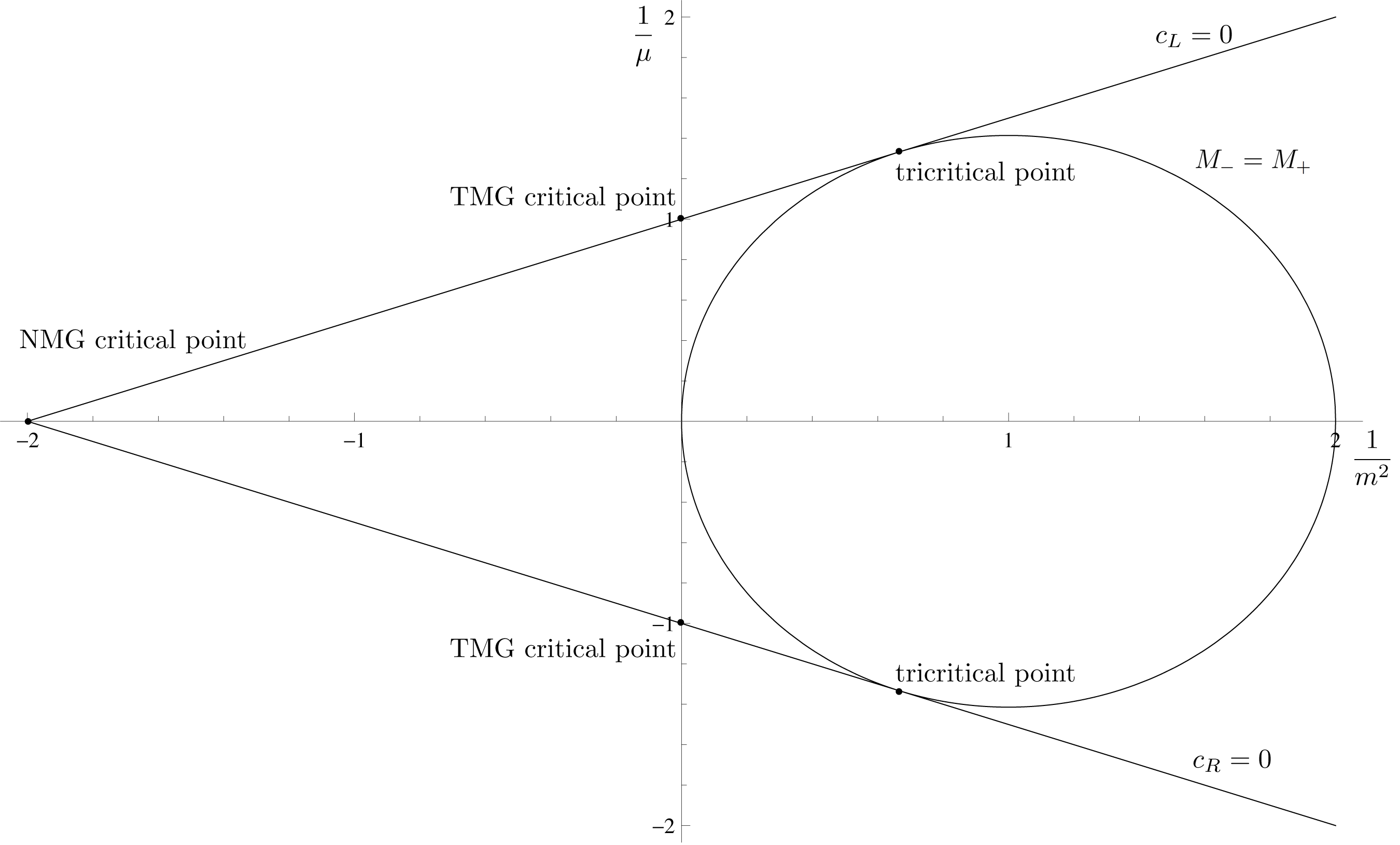}
\caption{\small The parameter space of GMG with $\ell= 1$ and $\sigma = +1$. In addition to the tricritical points in the left-
and right-moving sector the NMG and TMG critical points are displayed. For $\sigma = -1$ the plot looks the same, only mirrored
in the $1/\mu$-axis.}
\label{parameterspace_GMG}
\end{figure}

The action of General Massive Gravity (GMG) is given by \cite{Bergshoeff:2009aq}
\begin{equation} \label{SGMG}
S = \frac{1}{16 \pi G_N } \int \rmd^3 x \sqrt{-g} \left\{ \sigma R - 2 \Lambda_0 + \frac{1}{m^2} \left( R^{\mu \nu} R_{\mu \nu} - \frac38 R^2 \right) + \frac{1}{\mu} \mathcal{L}_{\mathrm{LCS}} \right\} \,,
\end{equation}
where the Lorentz--Chern--Simons term (LCS) is given by \cite{Deser:1981wh}
\begin{equation} \label{LCS}
\mathcal{L}_{\mathrm{LCS}} = \frac12 \varepsilon^{\mu \nu \rho} \left[\Gamma^\alpha_{\mu \beta} \partial_\nu \Gamma^\beta_{\rho \alpha} + \frac23 \Gamma^\alpha_{\mu \gamma} \Gamma^\gamma_{\nu \beta} \Gamma^\beta_{\rho \alpha}  \right] \,.
\end{equation}
The parameters $m$, $\mu$ are mass parameters, while $\sigma$ is a dimensionless sign parameter that takes on the values
$\pm 1$ and $\Lambda_0$ is the cosmological constant. In particular
this model has an AdS solution with cosmological constant $\Lambda = -1/\ell^2$ for
\begin{equation}
\Lambda_0 = \frac{1}{4 m^2 \ell^2} - \frac{\sigma}{\ell^2} \,.
\end{equation}
The linearized equations of motion of the GMG model are given by \cite{Liu:2009pha}
\begin{equation} \label{lineomGMG}
\left(\mathcal{D}^L \mathcal{D}^R \mathcal{D}^{M_+} \mathcal{D}^{M_-} h  \right)_{\mu \nu} = 0 \,,
\end{equation}
where $h_{\mu \nu}$ denotes the perturbation of the metric around a background spacetime, that will be taken as AdS in
the following. The differential operators $\mathcal{D}^M$, $\mathcal{D}^L$, $\mathcal{D}^R$ are given by
\begin{equation} \label{defdiffs}
(\mathcal{D}^M)_\mu{}^\beta = \delta_\mu^\beta + \frac{1}{M} \varepsilon_\mu{}^{\alpha \beta}
 \nabla_\alpha \,, \qquad
 (\mathcal{D}^{L/R})_\mu{}^\beta = \delta_\mu^\beta \pm \ell\varepsilon_\mu{}^{\alpha \beta} \nabla_\alpha \,.
\end{equation}
The mass parameters $M_\pm$ appearing in \eqref{lineomGMG} can be expressed in terms of the GMG parameters as follows:\footnote{
The mass parameters $M_\pm$ in eq.~\eqref{124} can assume both positive and negative values. The physical masses are thus given by the
absolute values of $M_\pm$. The helicities of the corresponding modes are then given by the signs of $M_\pm$. Note that $M_\pm$ do not
necessarily need to have opposite signs or helicities.}
\begin{equation}\label{124}
 M_+ = \frac{m^2}{2 \mu} + \sqrt{\frac{1}{2\ell^2} - \sigma m^2 + \frac{m^4}{4 \mu^2}} \,,
 \qquad M_- = \frac{m^2}{2 \mu} - \sqrt{\frac{1}{2\ell^2} - \sigma m^2 + \frac{m^4}{4 \mu^2}} \,.
\end{equation}
The GMG model has various critical points and lines in its parameter space where several of the differential operators
in \eqref{lineomGMG} degenerate. These were discussed in more detail in \cite{Grumiller:2010tj}. Figure \ref{parameterspace_GMG}
shows a plot of the parameter space of GMG. Here the critical lines where $c_L=0$ and $c_R =0$, whose expressions are given by \eqref{cLRGMG}, are displayed as well as the
critical curve where $M_+=M_-$. The NMG and TMG limits of GMG are on the $1/m^2$ and $1/\mu$-axis respectively and whenever
a critical line intersects with one of them, critical TMG or NMG is recovered. At the origin both masses become infinite and decouple.
This point corresponds to Einstein gravity in three dimensions. In the following we will be mainly interested in the tricritical
points, where three operators of \eqref{lineomGMG} degenerate. There are two such tricritical points, given by the following
parameter values:
\begin{align}
 \mathrm{point} \ 1 \ \ &: \ \ m^2\ell^2 = 2 \mu\ell = \frac32 \sigma \,, \label{DLDLDL}\\
 \mathrm{point} \ 2 \ \ &: \ \ m^2\ell^2 = -2 \mu\ell = \frac32 \sigma \,.
\end{align}
At point 1, the operators $\mathcal{D}^{M_+}$ and $\mathcal{D}^{M_-}$ degenerate with $\mathcal{D}^L$, whereas at point
2, they degenerate with $\mathcal{D}^R$. We will mainly focus on the first of these two critical points;
results for the second critical point are obtained in a similar manner and mainly follow from exchanging $L$ and $R$.

\subsection{Tricritical GMG as a log CFT}

According to the AdS/CFT correspondence, GMG around an AdS background is dual to a two-dimensional conformal field theory
(CFT), living on the boundary of AdS. The central charges of this CFT have been calculated in \cite{Liu:2009pha} and are given by
\begin{equation}\label{cLRGMG}
c_L = \frac{3\ell}{2 G_N} \left( \sigma + \frac{1}{2m^2\ell^2} - \frac{1}{\mu\ell} \right) \,,
  \qquad c_R = \frac{3\ell}{2 G_N} \left( \sigma + \frac{1}{2m^2\ell^2} + \frac{1}{\mu\ell} \right) \,.
\end{equation}
From \eqref{DLDLDL}, we then see that at the tricritical points the left- and right-moving central charges assume
the following values:
\begin{align}\begin{split}
 \mathrm{point} \ 1 \ \ :& \ \ c_L = 0 \,, \quad c_R = \frac{4\ell\sigma}{G_N} \,, \\
 \mathrm{point} \ 2 \ \ :& \ \  c_L = \frac{4 \ell\sigma}{G_N}\,, \quad c_R = 0 \,.
\end{split}\end{align}
We thus see that at the tricritical points the central charge of the sector where the degeneracy in \eqref{lineomGMG}
takes place, is zero. 

One can say more about the structure of the dual CFT, by examining the solutions to the linearized equations of motion
\eqref{lineomGMG} at the critical point and applying holographic reasoning. We will focus on the critical point 1, where
the equations of motion are given by
\begin{equation}
\left(\mathcal{D}^L \mathcal{D}^L \mathcal{D}^L \mathcal{D}^R h\right)_{\mu \nu} = 0 \,.
\end{equation}
One can show that the solution space of these equations is spanned by solutions $h^R$, $h^L$, $h^{\mathrm{log}}$ and
$ h^{\mathrm{log}^2}$ that obey
\begin{eqnarray}
& & (\mathcal{D}^R h^R)_{\mu \nu} = 0 \,, \nonumber \\
& & (\mathcal{D}^L h^L)_{\mu \nu} = 0 \,, \nonumber \\
& & (\mathcal{D}^L \mathcal{D}^L h^\mathrm{log})_{\mu \nu} = 0 \,, \qquad \mathrm{but} \ \ (\mathcal{D}^L h^{\mathrm{log}})_{\mu \nu} \neq 0 \,, \nonumber \\
& & (\mathcal{D}^L \mathcal{D}^L \mathcal{D}^L h^{\mathrm{log}^2})_{\mu \nu} = 0 \,, \qquad \mathrm{but} \ \ (\mathcal{D}^L\mathcal{D}^L h^{\mathrm{log}^2})_{\mu \nu} \neq 0 \,.
\end{eqnarray}
Using the explicit expressions of these modes \cite{Li:2008dq,Grumiller:2008qz}, one can see that the modes $h^R$, $h^L$
fall off towards the AdS boundary in the way that is given by the Brown--Henneaux boundary conditions.
The modes $h^{\mathrm{log}}$, $h^{\mathrm{log}^2}$ on the other hand do not obey the usual Brown--Henneaux boundary conditions,
but are characterized by a $\log$-, resp.~$\log^2$-asymptotic behavior towards the boundary. In formulating the
$\mathrm{AdS}_3/\mathrm{CFT}_2$ correspondence, the issue of boundary conditions is essential. In order to define a
theory of quantum gravity on $\mathrm{AdS}_3$, one has to specify boundary conditions, that are relaxed enough to allow
for finite mass excitations and restricted enough to allow for a well-defined action of the asymptotic symmetry group. In the
case of tricritical GMG, it has been shown in \cite{Liu:2009pha} that one can formulate a consistent set of boundary
conditions that allows for excitations with both $\log$ and $\log^2$ fall-off behavior towards the boundary.

The conserved Abbott--Deser--Tekin charges for GMG were also calculated in \cite{Liu:2009pha}. We are only interested in the charges of
tricritical GMG, using \eqref{DLDLDL}. For Brown--Henneaux boundary conditions we find
\begin{align}
Q^{\rm GMG}_L = 0 \,, \qquad {\rm and}\qquad Q^{\rm GMG}_R &= \frac{\sigma}{3\pi G_N\ell}\int\extd\phi\,\epsilon^-f_{--}\,,
\end{align}
equivalent to \eqref{QLBH} and \eqref{QRBH} resp.~for the PET model. Imposing log boundary conditions we find
\begin{align}
 Q^{\rm GMG}_L = 0 \,, \qquad {\rm and}\qquad Q^{\rm GMG}_R &= \frac{\sigma}{3\pi G_N\ell}\int\extd\phi\,\epsilon^-f^{\rm log}_{--}\,.
\end{align}
For $\log^2$  boundary conditions we obtain
\begin{align}
 Q^{\rm GMG}_L = \frac{\sigma}{6\pi G_N\ell}\int\extd\phi\,\epsilon^+f^{\rm log^2}_{++}\,, \qquad {\rm and}\qquad
 Q^{\rm GMG}_R &= \frac{\sigma}{3\pi G_N\ell}\int\extd\phi\,\epsilon^-f^{\rm log^2}_{--}\,.
\end{align}
Truncating tricritical GMG means that we restrict the theory to the $Q_L = 0$ sub-sector. This restriction reduces the
$\log^2$ boundary conditions to $\log$ boundary conditions. Charge conservation then guarantees that the boundary conditions
are preserved under time evolution and the sub-sector thus decouples from the full theory. 

As is the case in chiral gravity, the $Q_L = 0$ sub-sector still contains the right-moving massless Einstein modes.
However, in contrast to chiral gravity, the left-moving sector contains a non-trivial two-point function for the dual
logarithmic operators. From \cite{Grumiller:2010tj} we find
\begin{equation}
\vev{\cO^{\log}(z) \cO^{\log}(0)} = \frac{B_L}{2 z^4}\,,
\end{equation}
with $B_L = 4\ell \sigma / G_N$. Like in truncated PET gravity, this correlation function has the same structure as a
CFT two-point function.

\section{Details on the calculation of the boundary stress tensor}\label{stresstensorapp}

The variational principle for a higher-derivative theory, such as the PET model, is an ambiguous task.
Imposing boundary conditions
only on the metric field is not sufficient to make the whole boundary term vanish. To fix the boundary
conditions we will employ a method
put forward in \cite{Hohm:2010jc}, introducing auxiliary fields. The boundary conditions are then
given by demanding that the variations of the metric and all auxiliary fields vanish. The first variation of
the action is zero if the boundary term only consists of terms multiplying $\delta g_{\mu\nu}$,
$\delta f_{\mu\nu}$ and $\delta \lambda_{\mu\nu}$, but no derivatives thereof.
For the metric $g_{\mu\nu}$ and the auxiliary field $f_{\mu\nu}$ this implies adding a generalized
Gibbons--Hawking term \cite{Hohm:2010jc}. Unfortunately, our action depends also on explicit derivatives
of $\lambda_{\mu\nu}$, thus we have to add another counterterm to remove all boundary terms proportional
to $\nabla_\alpha\delta\lambda_{\mu\nu}$.

The boundary term of the variation of the action \eqref{aux2} in full detail reads
\begin{align}\begin{split}\label{thereallynotsoniceformula}
 \delta S &= \frac{1}{16 \pi G}\int_{\partial M}\extd^2x\,\sqrt{-\gamma}\Big\{
         -\big(A_y^{\phantom{y}y}\gamma^{ij}-A^{ij}\big)\delta K_{ij}+\big(A_y^{\phantom{y}y}K^{ij}
         -\nabla_k\big[\frac{y}{2}A_y^{\phantom{y}k})\big]\gamma^{ij}+ \\
     & \qquad+ y\nabla^iA_y^{\phantom{y}j}-\frac{y}{2}\nabla_\alpha A_y^{\phantom{y}\alpha}\gamma^{ij}
         -\frac{y}{2}\nabla_yA^{ij}\big)\delta\gamma_{ij} \\
     & \quad+ b_2\,n^\alpha\Big[\nabla_\alpha\lambda^{\mu\nu}\delta\lambda_{\mu\nu}
         -\lambda^{\mu\nu}\nabla_\alpha\delta\lambda_{\mu\nu}+
      \lambda g^{\mu\nu}\nabla_\alpha\delta\lambda_{\mu\nu}-(\nabla_\alpha\lambda)g^{\mu\nu}\delta\lambda_{\mu\nu}+\\
   &\qquad+2\lambda_\alpha^{\phantom{\alpha}\nu}\nabla^{\delta}\delta\lambda_{\delta\nu}
         -2(\nabla^\nu\lambda_\nu^{\phantom{\nu}\delta})\delta\lambda_{\alpha\delta}
         -2\lambda\nabla^\nu\delta\lambda_{\alpha\nu}+2(\nabla^\nu\lambda)\delta\lambda_{\alpha\nu}+ \\
     &\qquad+\lambda^{\mu\nu}(\nabla^\delta\lambda_{\mu\nu})\delta g_{\delta\alpha}
         -\frac{1}{2}\lambda^{\mu\nu}(\nabla_\alpha\lambda_{\mu\nu})g^{\delta\beta}\delta g_{\delta\beta}
         +2\lambda^{\mu\nu}(\nabla_\alpha\lambda_\nu^{\phantom{\nu}\delta})\delta g_{\delta\mu}- \\
     &\qquad-2\lambda^{\mu\nu}(\nabla^\delta\lambda_{\alpha\nu})\delta g_{\delta\mu}
         +\lambda^{\mu\nu}\lambda_\nu^{\phantom{\nu}\delta}\nabla_\alpha\delta g_{\delta\mu}
         -\nabla_\alpha(\lambda^{\mu\nu}\lambda_\nu^{\phantom{\nu}\delta})\delta g_{\delta\mu}+ \\
     &\qquad+2\lambda_\alpha^{\phantom{\alpha}\nu}(\nabla^\mu\lambda_\nu^{\phantom{\nu}\delta})\delta g_{\delta\mu}
         -\lambda(\nabla^\mu\lambda)\delta g_{\mu\alpha}+\frac{1}{2}\lambda(\nabla_\alpha\lambda)g^{\mu\nu}\delta g_{\mu\nu}- \\
     &\qquad-2\lambda(\nabla_\alpha\lambda^{\delta\nu})\delta g_{\delta\nu}
         -\lambda\lambda^{\delta\nu}\nabla_\alpha g_{\delta\nu}+\nabla_\alpha(\lambda\lambda^{\delta\nu})\delta g_{\delta\nu}
         -3\lambda_\alpha^{\phantom{\alpha}\nu}(\nabla^\delta\lambda_\nu^{\phantom{\nu}\beta})\delta g_{\delta\beta}\\
     &\qquad+\lambda^{\beta\nu}(\nabla^\delta\lambda_{\alpha\nu})\delta g_{\beta\delta}
         -\lambda^{\delta\nu}(\nabla_\alpha\lambda_\nu^{\phantom{\nu}\beta})\delta g_{\delta\beta}
         -4\lambda^{\mu\nu}(\nabla_\mu\lambda_\nu^{\phantom{\nu}\delta})\delta g_{\alpha\delta} +\\
     &\qquad+2\lambda^{\delta\nu}(\nabla_\delta\lambda_{\alpha\nu})g^{\mu\beta}\delta g_{\mu\beta}
         -4\lambda_{\alpha\nu}\lambda^{\nu\delta}\nabla^\beta\delta g_{\delta\beta}
         +2\lambda_{\alpha\nu}\lambda^{\nu\delta}g^{\mu\beta}\nabla_\delta\delta g_{\mu\beta} +\\
     &\qquad +4\nabla^\mu(\lambda_{\mu\nu}\lambda^{\nu\delta})\delta g_{\delta\alpha}
         -2\nabla^\delta(\lambda_{\delta\nu}\lambda_\alpha^{\phantom{\alpha}\nu})g^{\mu\beta}\delta g_{\mu\beta}
         -2\nabla_\alpha(\lambda\lambda^{\delta\beta})\delta g_{\delta\beta} +\\
     &\qquad+2\lambda(\nabla^\delta\lambda_\alpha^{\phantom{\alpha}\beta})\delta g_{\delta\beta}
         -\lambda(\nabla_\alpha\lambda^{\delta\beta})\delta g_{\delta\beta}
         +8\lambda(\nabla^\beta\lambda_\beta^{\phantom{\beta}\delta})\delta g_{\delta\alpha} -\\
     &\qquad-4\lambda(\nabla^\delta\lambda_{\delta\alpha})g^{\mu\beta}\delta g_{\mu\beta}
         +2\lambda\lambda^{\delta\beta}\nabla_\alpha\delta g_{\delta\beta}\Big]
\Big\}\,,
\end{split}\end{align}
where $n^{\mu}$ is the vector normal to the boundary, $K_{ij}$ the extrinsic curvature and
$A_{\mu\nu}=(\sigma-f/2) g_{\mu\nu}+f_{\mu\nu}$. In the first line we used the identities of our background \eqref{poincaremetric}
\begin{align}\begin{split}
 n^\mu=[0,0,-y]\qquad K_{ij}=\nabla_{(i}n_{j)}\qquad\delta K_{ij}=-\frac{y}{2} \partial_y\delta\gamma_{ij} \\
  \Gamma^y_{ij}=yK_{ij}\qquad\Gamma^i_{yk}=-\frac{1}{y}K^i_k \qquad \Gamma^i_{yy}=\Gamma^y_{yi}=0 \,.
\end{split}\end{align}
To get rid of all derivatives on $\delta\lambda_{\mu\nu}$ we add the following term:
\begin{align}\label{auxGGH}
 I_{\rm aux-\lambda}= \frac{b_2}{16 \pi G}\int_{\partial M}\extd^2x \sqrt{-\gamma}\,n^\alpha\big[
              (\lambda^{\mu\nu}-\lambda g^{\mu\nu})\nabla_\alpha\lambda_{\mu\nu}
             +2\lambda\nabla^\nu\lambda_{\alpha\nu}
             -2\lambda_\alpha^{\phantom{\alpha}\mu}\nabla^\nu\lambda_{\mu\nu}\big] \,.
\end{align}
This also removes many terms of the form $\nabla_\mu\delta g_{\beta\nu}$, but not all. 
Varying the term \eqref{auxGGH} introduces further derivatives of variations of the metric,
hence we can only fix the generalized Gibbons--Hawking term
after taking into account all contributions from \eqref{auxGGH}. After performing some algebra we see
that we have to add
\begin{align}\label{GGHterm}
 I_{\rm GGH}=\frac{1}{16 \pi G}\int_{\partial M}\extd^2x \sqrt{-\gamma}\,\big[
              A_y^{\phantom{y}y}K+A^{ij}K_{ij}-2b_2\lambda\lambda_y^{\phantom{y}y}K-2b_2\lambda\lambda^{ij}K_{ij}\big]
\end{align}
as the generalized Gibbons--Hawking counterterm. We have now obtained a boundary term that schematically takes the form
\begin{align}
 \delta(S+I_{\rm aux-\lambda}+I_{\rm GGH})\big|_{\partial M}= \int_{\partial M} \extd^2x \sqrt{-\gamma}
      \Big\{\big(\ldots\big)\delta\gamma_{ij}+\big(\ldots\big)\delta f_{ij}+\big(\ldots\big)\delta\lambda_{\mu\nu}\Big\}\,.
\end{align}
Thus, the first variation of the action vanishes if we set the variations $\delta\gamma_{ij}$, $\delta f_{ij}$
and $\delta\lambda_{\mu\nu}$ to zero at the boundary. 

Finally, we need another holographic counterterm
\begin{align}\label{constcounterterm}
 I_{\rm ct}=-\frac{1}{16 \pi G}\int_{\partial M}\extd^2x \sqrt{-\gamma}\,\Big(2\sigma+\frac{\beta}{\ell^2}-\frac{2b_2}{\ell^4}\Big)
\end{align}
to find the renormalized action. Plugging in the on-shell values for $f_{\mu\nu}$ and $\lambda_{\mu\nu}$
we can read off the stress tensor from the identity
\begin{align}\label{actren}
 \delta S_{\rm ren} =\delta(S+I_{\rm aux-\lambda}+I_{\rm GGH}+I_{\rm ct})=
          \frac{1}{2} \int_{\partial M}\extd^2x \sqrt{-\gamma}\,\,T^{ij}\delta\gamma_{ij}^{(0)} \,.
\end{align}
The final result obtained from \eqref{actren} is
\begin{align}
 16 \pi G\,T^{\rm 3critgrav}_{ij}=\Big(2\sigma+\frac{\beta}{\ell^2}+\frac{3b_2}{\ell^4}\Big)\gamma_{ij}^{(2)}
                 -\Big(2\sigma+\frac{\beta}{\ell^2}\Big)\gamma_{ij}^{(0)}\gamma_{kl}^{(2)}\gamma^{kl}_{(0)} \,.
\end{align}

\section{Graviton energies}\label{Energyapp}
This appendix is devoted to the calculation of the on-shell energies of the linearized graviton modes given in section
\ref{PETCFTsection}. We may do so by constructing the Hamiltonian of the six-derivative gravity theory. In order to achieve this
we need to define what are the canonical variables. This is most easily done by using the auxiliary field formulation of section
\ref{auxfieldsssec}, where the definition of the canonical variables and the Hamiltonian is the standard one.

\subsection{Energy of the massless and the massive modes}
For the on-shell energies of the massless and the massive modes, we may use the diagonalized Lagrangian \eqref{L2diag}. The original
metric perturbation $h_{\mu\nu}$, defined as $g_{\mu\nu} = \bg_{\mu\nu} + h_{\mu\nu}$, consists of the massless mode and both
massive modes. In the renewed definition, $h'_{\mu\nu}$ is only the massless mode. The massive modes $\psi^{(M_+)}_{\mu\nu}$ and
$\psi^{(M_-)}_{\mu\nu}$ are proportional to $k'_1{}_{\mu\nu}$ and $k'_2{}_{\mu\nu}$, resp. We will take the constant of proportionality
as the inverse of the shifts in \eqref{shift1}:
\begin{equation}
k'_1{}_{\mu\nu} = \frac{\bar{\sigma}}{2b_2 M_-^2}\psi^{M_+}_{\mu\nu} \,, \qquad 
k'_2{}_{\mu\nu} = \bar{\sigma} \psi^{M_-}_{\mu\nu} \,,
\end{equation}
where $\psi^{M_{\pm}}$ are the solutions \eqref{psisol} with the weights \eqref{hleft} and \eqref{hright}. We may now calculate the
Hamiltonian from \eqref{L2diag}, using as canonical variables $h'_{\mu\nu}, k'_1{}_{\mu\nu}$ and $k'_2{}_{\mu\nu}$. Since we are only
interested in the on-shell energies, we may take the fields to be transverse and traceless when computing the Hamiltonian. The result is:
\begin{align}
H = & \; \frac{1}{32 \pi G} \int d^2x \sqrt{-\bg} \bigg\{ 
- \bar{\sigma} \bnabla^0 h'{}^{\mu\nu} \dot{h'}_{\mu\nu} + \frac{4b_2}{\bar{\sigma}}\left(\bar{\sigma} + b_2M_-^4 \right)  \bnabla^0 k'_1{}^{\mu\nu} \dot{k'}_{1\mu\nu} \\
& \; + \frac{1}{\bar{\sigma}^2}\left( \bar{\sigma} + b_2M_-^4 \right) \bnabla^0 k'_2{}^{\mu\nu} \dot{k'}_{2\mu\nu} - \cL^{(2)} \bigg\}  \,, \nonumber
\end{align}
with $\cL^{(2)}$ given by \eqref{L2diag}. Evaluating this for each of the modes we obtain the expression for the on-shell energies of the linearized modes:
\begin{align}\label{Emassless}
E^{0} & = - \frac{\bar{\sigma}}{32 \pi G} \int d^2 x \sqrt{-\bg}  \bnabla^0\psi^{0}{}^{\mu\nu} \dot{\psi}_{\mu\nu}^{0}  \\
E^{M_{\pm}} & =  \frac{1}{32 \pi G}\left(\bar{\sigma} + b_2 M_{\pm}^4 \right) \int d^3 x \sqrt{-\bg} \bnabla^0\psi^{M_{\pm}}{}^{\mu\nu} \dot{\psi}_{\mu\nu}^{M_{\pm}} \,. \label{Emassive}
\end{align}
When taking the NMG limit $b_2 \to 0$, one can see from \eqref{Masses} that of the masses becomes infinite and decouples. For the other massive mode and the massless mode the energy reduces to the expressions found in \cite{Liu:2009bk}. On the critical line \eqref{critline} and at the tricritical point \eqref{tricritpoint} we can not trust the expressions above, since the Lagrangian from which they are derived is no longer valid. Instead, we must use the Lagrangians \eqref{L2critline} and \eqref{L2tricrit} to define the Hamiltonian at these special values in the parameter space. We will do so below. 

In order to make sure that there are no ghosts, \eqref{Emassless} and \eqref{Emassive} must have the same sign. The integrals are all negative, so we must constrain $\bar{\sigma} \geq 0$ and $(\bar{\sigma} + b_2 M_{\pm}^4) \leq 0$. It can be shown that this constraint is equivalent to demanding that the kinetic terms in \eqref{L2diag} all have the same sign. 

\subsection{Energy of the log and log$^2$ modes}
At the critical line \eqref{critline} the linearized Lagrangian becomes \eqref{L2critline}. It consists of a Fierz-Pauli Lagrangian for one massive spin-2 field and a part which resembles the linearized Lagrangian of critical NMG. From the equations of motion of \eqref{L2critline} we may conclude that $k''_{1\mu\nu}$ corresponds to the remaining massive mode $\psi_{\mu\nu}^{M'}$. The field $h''_{\mu\nu}$ now corresponds to the log mode. If we take $h''_{\mu\nu} = \psi^{\log}_{\mu\nu}$, then from the equations of motion one can derive that
\begin{equation}\label{k2ismassless}
k''_{2\mu\nu} = 2 \left(2\sigma + \Lambda^2 b_2\right) \psi^{0}_{\mu\nu}\,.
\end{equation}
The Hamiltonian, constructed from \eqref{L2critline} with the double primed fields as canonical variables and again taking the fields to be transverse and traceless, reads
\begin{equation}\label{Hcritline}
H = \; \frac{1}{16 \pi G} \int d^2x \sqrt{-\bg} \bigg\{ 
 \bnabla^0 k''_2{}^{\mu\nu} \dot{h}''_{\mu\nu} +  2b_2 \bnabla^0 k''_1{}^{\mu\nu} \dot{k}''_{1\mu\nu} - \cL^{(2)} \bigg\} \,,
\end{equation}
where here $\cL^{(2)}$ is given by \eqref{L2critline}. From this we can derive that the energy of the massive mode is given by \eqref{Emassive} with $\bar{\sigma} =0$ and $M_{\pm} = M'$, while the massless mode has zero energy. To obtain the energy for the log mode, we plug in $h_{\mu\nu} = \psi_{\mu\nu}^{\log}$ and \eqref{k2ismassless}:
\begin{equation}\label{Elog}
E^{\log}  = \frac{1}{8 \pi G} \left(2 \sigma + \Lambda^2 b_2 \right)   \int d^2 x \sqrt{-\bg}  \bnabla^0\psi^{0}{}^{\mu\nu} \dot{\psi}_{\mu\nu}^{\log} \,.
\end{equation}
At the tricritical point this expression is ill-defined, since the Lagrangian \eqref{L2critline} is not valid at this point. At this point the linearized Lagrangian \eqref{linShk} reduces to \eqref{L2tricrit}. For transverse and traceless fields the corresponding Hamiltonian is given by \eqref{Hcritline} with the primes removed and with $\cL^{(2)}$ given by equation \eqref{L2tricrit}. The equations of motion \eqref{w4}-\eqref{w6} tell us that if we take $h_{\mu\nu} = \psi^{\log^2}_{\mu\nu}$, then $k_{1\mu\nu} = \Lambda \psi^{\log}_{\mu\nu} + \frac12 \Lambda \psi_{\mu\nu}^{0}$ and $k_{2\mu\nu} = 4b_2 \Lambda^2 \psi_{\mu\nu}^{0}$. This gives at the tricritical point
\begin{align}\label{Elog2}
E^{\log^2}  = - \frac{\sigma}{4 \pi G}  \int d^2 x \sqrt{-\bg} \Big\{ & 2\bnabla^0\psi^{0}{}^{\mu\nu} \dot{\psi}_{\mu\nu}^{\log^2} + \bnabla^0\psi^{\log}{}^{\mu\nu} \dot{\psi}_{\mu\nu}^{\log} \\
&  + \bnabla^0\psi^{0}{}^{\mu\nu} \dot{\psi}_{\mu\nu}^{\log} + \frac14 \bnabla^0\psi^{0}{}^{\mu\nu} \dot{\psi}_{\mu\nu}^{0}  \Big\} \,. & \nonumber
\end{align}
This expression is finite everywhere and positive for $\sigma = +1 $. The on-shell energy of the massless mode may be calculated by taking $h_{\mu\nu} = \psi_{\mu\nu}^{0}$. This choice requires that $k_{1\mu\nu}$ and $k_{2\mu\nu}$ are zero and the on-shell energy of the massless mode vanishes. For the log mode we take $h_{\mu\nu} = \psi_{\mu\nu}^{\log}$ which implies that $k_{1\mu\nu} = \Lambda \psi_{\mu\nu}^{0}$ and $k_{2\mu\nu} = 0$. This leads to
\begin{equation}\label{Elogtricrit}
E^{\log}_{\rm{tricritical}}  = - \frac{\sigma}{4 \pi G}  \int d^2 x \sqrt{-\bg}  \bnabla^0\psi^{0}{}^{\mu\nu} \dot{\psi}_{\mu\nu}^{0} \,.
\end{equation}
Even though the truncation of the log$^2$ modes is defined via boundary conditions for the full non-linear theory, at the linearized level the non-vanishing of the log mode energy at the tricritical point suggests that the truncated theory may be non-trivial. 

\providecommand{\href}[2]{#2}\begingroup\raggedright\endgroup


\end{document}